\documentclass[aps,prx,showpacs,preprintnumbers,twocolumn,superscriptaddress,longbibliography]{revtex4-2}

\usepackage[T1]{fontenc}
\usepackage{newtxtext}
\usepackage{newtxmath}

\usepackage{amsmath}
\usepackage{mathtools}
\usepackage{amsfonts}
\usepackage{braket}

\newcommand{\e}{\mathrm{e}}

\newcommand{\beq}{\begin{equation}}
\newcommand{\eeq}{\end{equation}}

\newcommand{\ignore}[1]{}
\DeclareMathOperator{\tr}{tr}

\usepackage{color}
\usepackage[dvipsnames]{xcolor}
\usepackage{colortbl}
\usepackage[colorlinks,citecolor=blue,linkcolor=blue,urlcolor=blue]{hyperref}

\definecolor{darkgreen}{rgb}{0.55, 0.71, 0.00}
\definecolor{Gray}{gray}{0.9}

\usepackage{graphicx}
\usepackage[all]{hypcap}

\begin{document}

\title{{Quantum many-body Jarzynski equality and dissipative noise on a digital quantum computer}}

\author{Dominik Hahn}
\email{hahn@pks.mpg.de}
\affiliation{Max Planck Institute for the Physics of Complex Systems, N\"{o}thnitzer Str.~38, 01187 Dresden, Germany}

\author{Maxime Dupont}
\affiliation{Department of Physics, University of California, Berkeley, California 94720, USA}
\affiliation{Materials Sciences Division, Lawrence Berkeley National Laboratory, Berkeley, California 94720, USA}

\author{Markus Schmitt}
\affiliation{Institute for Theoretical Physics, University of Cologne, 50937 Cologne, Germany}

\author{David J. Luitz}
\affiliation{Max Planck Institute for the Physics of Complex Systems, N\"{o}thnitzer Str.~38, 01187 Dresden, Germany}
\affiliation{Physikalisches Institut, Universit\"at Bonn, Nussallee 12, 53115 Bonn, Germany}

\author{Marin Bukov}
\email{mgbukov@phys.uni-sofia.bg}
\affiliation{Max Planck Institute for the Physics of Complex Systems, N\"{o}thnitzer Str.~38, 01187 Dresden, Germany}
\affiliation{Department of Physics, St.~Kliment Ohridski University of Sofia, 5 James Bourchier Blvd, 1164 Sofia, Bulgaria}

\begin{abstract}
    The quantum Jarzynski equality and the Crooks relation are fundamental laws connecting equilibrium processes with nonequilibrium fluctuations. They are promising tools to benchmark quantum devices and measure free energy differences. 
    While they are well established theoretically and also experimental realizations for few-body systems already exist, their experimental validity in the quantum many-body regime has not been observed so far.
    Here, we present results for nonequilibrium protocols in systems with up to sixteen interacting degrees of freedom obtained on trapped ion and superconducting qubit quantum computers, which test the quantum Jarzynski equality and the Crooks relation in the many-body regime.
    To achieve this, we overcome present-day limitations in the preparation of thermal ensembles and in the measurement of work distributions on noisy intermediate-scale quantum devices. We discuss the accuracy to which the Jarzynski equality holds on different quantum computing platforms subject to platform-specific errors.
    The analysis reveals the validity of Jarzynski's equality in a regime with energy dissipation,  compensated for by a fast unitary drive. This provides new insights for analyzing errors in many-body quantum simulators.
\end{abstract}
\maketitle

\section{\label{sec:intro}Introduction}

More than a century after the foundations of thermodynamics and equilibrium statistical mechanics have been laid, the field of nonequilibrium physics remains an area of active research. 
The study of how macroscopic properties arise from microscopic quantum dynamics, initiated the development of quantum thermodynamics~\cite{Talkner2020Colloquium,Millen_2016Perspective,binder2018thermodynamics}. 
Significant theorerical breakthroughs include the widely-applicable eigenstate thermalization hypothesis~\cite{Deutsch1991Quantum,Srednicki1994thermalization,rigol2008thermalization,d2016quantum,Deutsch_2018Eigenstate}, and a handful of interesting cases of its violation~\cite{Nandkishore2015Many-Body,Abanin2019Colloquium,serbyn2021quantum}; at the same time, ongoing experimental progress made it possible to cool down highly controlled systems to temperatures dominated by quantum rather than thermal fluctuations~\cite{an2015experimental,blatt2012quantum,gross2017quantum,schafer2020tools}. Despite this progress, we still lack a comprehensive theoretical framework or a complete set of principles to describe macroscopic phenomena in nonequilibrium physics. 

\begin{figure}[t!]
	\centering
	\includegraphics[width=1\columnwidth]{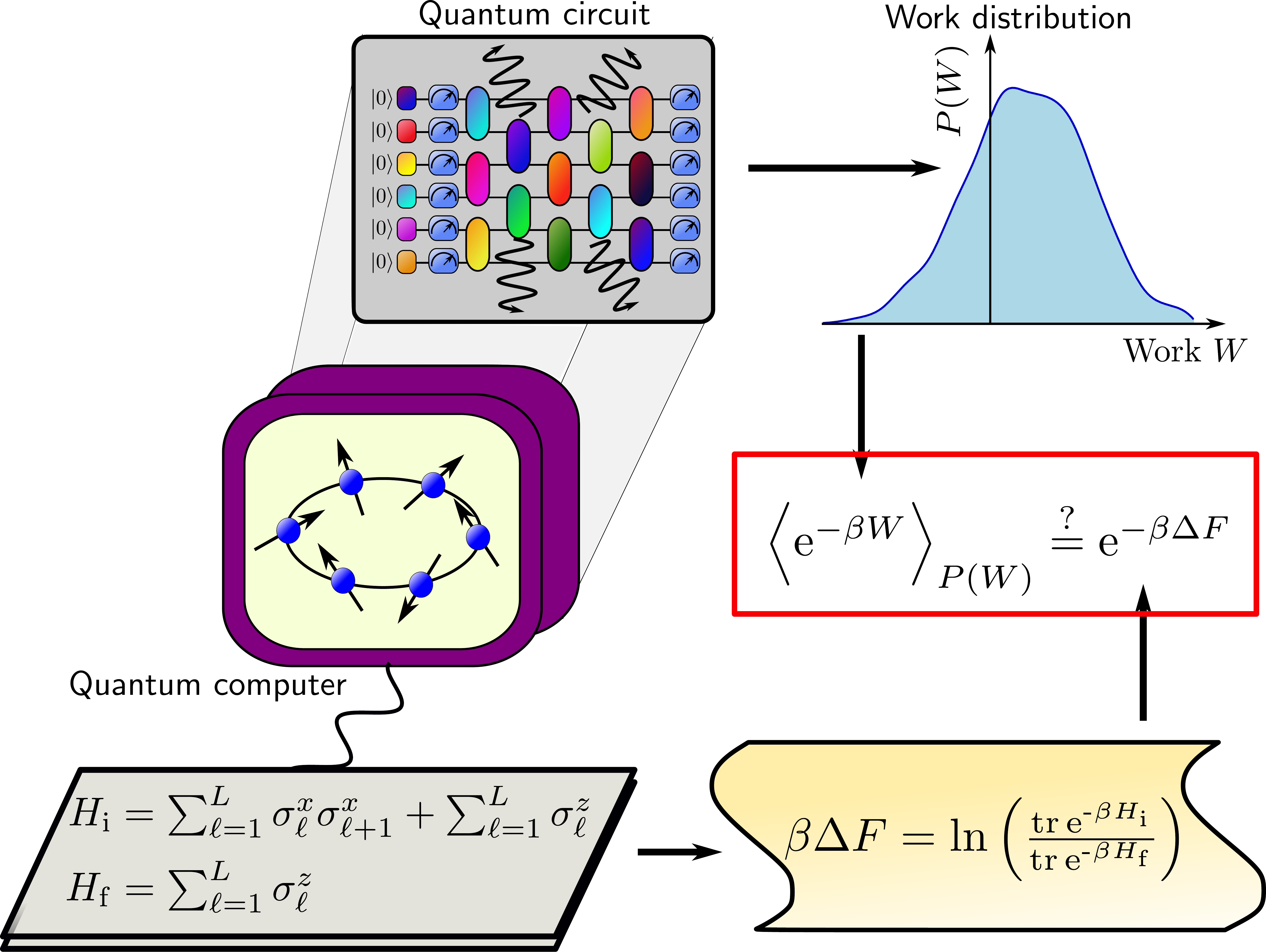}
	\caption{
	Schematics of our protocol to test the quantum Jarzynski equality, cf.~Eq.~\eqref{eq:Jarzynski}, in the quantum many-body regime. We simulate the dynamics of a system of interacting qubits initiated in a thermal state of the transverse-field Ising chain $H_{\mathrm{i}}$. The qubits then evolve under a non-equilibrium process on a quantum computer affected by energy dissipation. Finally, we extract the work distribution for the quantum circuit w.r.t.~the final Hamiltonian $H_{\mathrm{f}}$. 
	At the same time, we independently compute the exact theory prediction for the free energy difference. We compare both results against each other to test the validity of Jarzynski's equality [red box]. 
	}
	\label{fig:schematic}
\end{figure}

A remarkable achievement in the field is the Jarzynski equality~\cite{Jarzynski1997Nonequilibrium,broedersz2022twenty}. It establishes a mathematical relation between the 
work $W$ applied in a time-dependent process, 
and the free energy difference $\Delta F$ between the initial thermal ensemble and the thermal ensemble associated with the final Hamiltonian: 
\begin{equation}
    \Bigl\langle{\e^{-\beta W}}\Bigr\rangle_{P(W)} = \e^{- \beta \Delta F}~~\Longleftrightarrow~~\ln\Bigl\langle\e^{-\beta W_{\text{diss}}}\Bigr\rangle_{P(W)}=0,
\label{eq:Jarzynski}
\end{equation}
where $\beta=T^{-1}$ is the inverse temperature (we work in units $k_\mathrm{B}=1$, $\hbar=1$), and $W_\mathrm{diss}=W-\Delta F$ is the dissipated work done during the process; the average $\langle\cdot\rangle_{P(W)}$ is performed over the work distribution ${P(W)}$ obtained from repeated applications of the protocol.
Remarkably, Eq.~\eqref{eq:Jarzynski} holds for any nonequilibrium protocol without restrictions; for close-to-equilibrium processes it reduces to the well-known fluctuation-dissipation theorem~\cite{Jarzynski1997Nonequilibrium}.

The classical Jarzynski equality was verified in a number of experiments, ranging from stretching of single molecules~\cite{liphardt2002equilibrium,harris2007experimental} to mechanical systems~\cite{douarche2005experimental}, optical tweezers~\cite{blickle2006thermodynamics}, and electronic systems~\cite{saira2012test,Wimsatt2021Harnessing,Saira2020Nonequilibrium}.

For quantum systems, a difficulty has been identified with measuring work~\cite{Talkner2007Fluctuation,Aaberg2018FullyQuantum,Vinjanampathy_2016}. Nevertheless, Jarzynski's equality was generalized to closed quantum systems~\cite{kurchan2001quantum,tasaki2000jarzynski}, systems subject to dephasing~\cite{Mukamel2003Quantum}, general unital~\cite{Rastegin_2013Non-equilibrium} and stochastic~\cite{Rastegin2014Jarzynski} quantum maps, and systems with feedback-control~\cite{Sagawa2010Generalized,toyabe2010experimental}. Recently, there have been increasing efforts to extend the quantum Jarzynski equality to generic open systems using one-point-measurement schemes and the notion of "optimal guessed" quantum work~\cite{SoneQuantum2020,Beyer2021Workas, Deffner2016Quantumwork}. 

In the following, we focus on a formulation of the quantum Jarzynski equality valid for unital quantum channels, where work is defined by means of a two-point measurement scheme~\cite{kurchan2001quantum,tasaki2000jarzynski,Mukamel2003Quantum}: Let us denote the initial energy levels by $E^\mathrm{i}_m$, and final energy levels -- by $E^\mathrm{f}_n$. In the following, we refer to a projective measurement in an eigenstate of the initial or final Hamiltonian, as an energy measurement; the measurement output is the corresponding eigenenergy  $E^\mathrm{i}_m$ or $E^\mathrm{f}_n$, respectively. The work is then defined as the difference between the measured final and initial energies,
\begin{equation}
    W=E^\mathrm{f}_n - E^\mathrm{i}_m.
    \label{eq:work}
\end{equation}
While this definition does not generalize to the most general open systems, together with recent progress in quantum simulation, it allowed for the experimental test of Eq.~\eqref{eq:Jarzynski}~using  trapped ions~\cite{an2015experimental,smith2018verification}, cold atoms~\cite{cerisola2017using}, nuclear magnetic resonance (NMR) experiments~\cite{Bartalhao2014Experimental}, nitrogen-vacancy (NV) centers~\cite{hernandez2021experimental}, and superconducting qubits~\cite{Solfanelli2021Experimental}.

Besides its fundamental importance in quantum statistical physics, testing the quantum Jarzynski equality is also of practical interest, as it allows us to measure free energy differences, which can be used, e.g., to characterize the onset of chemical reactions. In a recent study, Jarzynski's equality was used to extract approximate free energy differences in two- or three-qubit systems using minimally entangled typical thermal states~\cite{bassman2021computing}.

In this work, we propose the idea  that the quantum Jarzynski equality using a two-point measurement scheme provides a valuable benchmark for the performance of quantum devices. The only requirement for the equality to hold is the doubly-stochastic property of the underlying dynamics (see Appendix~\ref{sec:theory} and~\cite{tasaki2000jarzynski,smith2018verification} for more details). This property is fulfilled for all unital channels, including pure unitary dynamics and dephasing noise. However, it does not hold for processes with energy dissipation. By experimentally testing Eq.~\eqref{eq:Jarzynski}, it is in principle possible to isolate contributions of processes violating double-stochasticity present during the dynamics and the measurement, which is of fundamental interest for improving current quantum technologies.

At the same time, a complete understanding of Jarzynski's equality in a quantum many-body setting is still lacking.
Previous verification experiments with quantum simulators were performed for single-, two-, and three-particle systems~\cite{an2015experimental,smith2018verification,cerisola2017using,Bartalhao2014Experimental,Solfanelli2021Experimental,bassman2021computing,hernandez2021experimental}. It is, therefore, crucial to close this gap and investigate systems of $L{\gtrsim}8$ spin-$1/2$ particles, where many-body characteristics begin to emerge~\cite{schiulaz2018few,zisling2021many}. 

In the many-body regime (i.e., for many interacting degrees of freedom), the work distribution broadens with the square root of the system size~\cite{rigol2008thermalization}, which requires an exponentially large number of projective measurements in order to estimate the left-hand-side of Eq.~\eqref{eq:Jarzynski}. 
Moreover, for quantum systems, further challenges arise: 
First, work fluctuations are not a measurable observable in quantum mechanics \cite{Talkner2007Fluctuation}. Hence, testing the quantum Jarzynski equality presumes the ability to measure in the energy eigenbasis. 
This is notoriously difficult in practice since many-body energy eigenstates are typically volume-law entangled in real space.
Second, preparing a quantum many-body system in a close-to-perfect thermal state can be demanding, and often comes with a substantial overhead of resource costs~\cite{Poulin2009Sampling,White2009Minimally}.
For these and related reasons, the quantum many-body regime provides formidable challenges.

In the present study, we use classical presampling or mid-circuit measurements to prepare thermal ensembles. In particular, mid-circuit measurements allow us to prepare a thermal ensemble of the transverse field Ising model~\cite{CerveraLierta2018exactisingmodel} for up to $L=16$ qubits. In contrast to many common approaches~\cite{Wu2019Variational,Poulin2009Sampling,White2009Minimally,Zhu2020Generation,Riera2012Thermalization}, this enables us to prepare the thermal initial state of the transverse field Ising model without the overhead of using ancilla qubits. This innovation allows us to reach system sizes one order of magnitude larger compared to previous studies, which opens the possibility of using the Jarzynski equality as a benchmark for many-body quantum devices.

In a simulation on digital quantum computers, we test Jarzynski's equality for up to $L=16$ qubits in a system of strongly interacting spins subject to a nonequilibrium protocol.
We also analyze the quantum Crooks relation~\cite{Crooks1999Entropy} [cf.~Eq.~\eqref{Crooks element log}] for $L=8$ qubits -- an infinitesimal version of Jarzynski's equality -- which had hitherto only been tested for a single two-level system~\cite{Bartalhao2014Experimental,Batalhao2015Irreversibility}.
Unlike earlier experiments~\cite{Bartalhao2014Experimental,an2015experimental,bassman2021computing,smith2018verification}, we do not use a parametric quench, but a protocol of random entangling gates.

Moreover, we compare the accuracy of our results with the relaxation times on the quantum processor, and identify a novel feature in this regime of nonequilibrium qubit dynamics: we show that Jarzynski's equality holds approximately even in the presence of accumulating dissipation effects, so long as the execution time of gates is short compared to the thermal qubit relaxation time  [so-called $T_1$]. In addition, we show that our nonequilibrium protocol improves the results in comparison to a pure energy dissipation process.

This paper is organized as follows: After introducing the challenges and their resolutions to test the quantum Jarzynski equality on a digital quantum computer in Sec.~\ref{sec:simulation}, we provide results for the Jarzynski equality and the Crooks relation in Sec.~\ref{sec:JarzynskiManybody}. In Sec.~\ref{sec:theory_analysis}, we develop a theory to understand the results in the presence of energy dissipation. Finally, we compare our results with previous experiments in Sec.~\ref{sec:Discussion} and discuss the potentials of our work for benchmarking quantum devices and measuring free energy differences.

\section{\label{sec:simulation}Simulation on Quantum Computers}

\begin{figure*}[t!]
	\centering
	\includegraphics[width=0.95\textwidth]{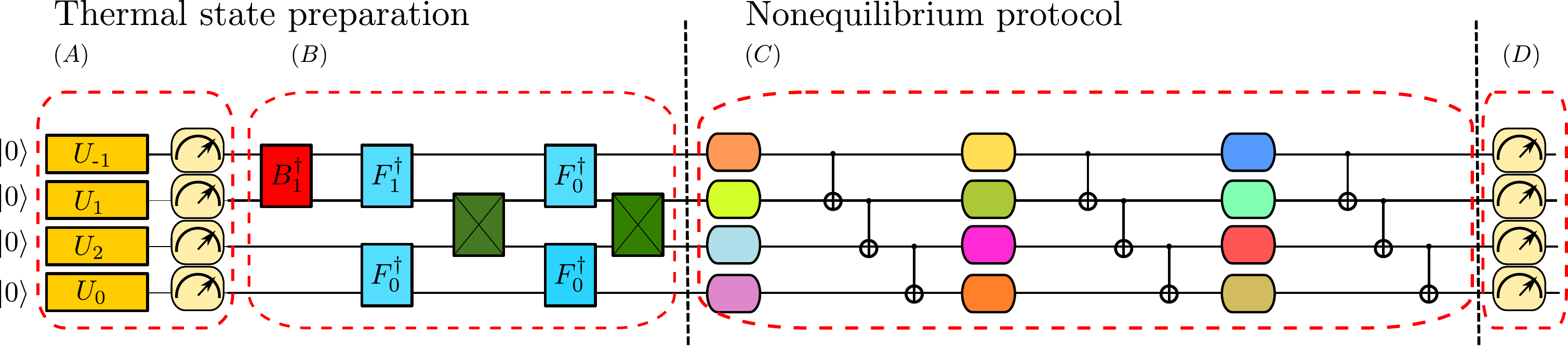}
	\caption{
	The protocol of our quantum experiments (A) The unitary rotation gate defined in Eq.~\eqref{rotation} and subsequent measurements prepare a thermal ensemble of the transverse field Ising model. The measurement has two effects: it prepares a thermal ensemble and is used as the initial energy measurement to determine the work distribution. (B) Transformation from the initial energy to the computational basis using a Bogoliubov~(red) and a Fourier transformation (light blue). The gates are defined in Appendix~\ref{Ising preparation}, the fermionic SWAP operators~(dark green) keep track of the correct sign structure under permutation.
	(C) Now we apply the actual nonequilibrium protocol. In our case we execute a "random" circuit with $k=3$ blocks. Each block consists of a layer of single qubit Haar random unitary gates~(different colors) and a sequential CNOT layer. (D) Final measurement of the qubits in the computational basis corresponding to the final energy measurement of of Eq.~\eqref{Final Hamiltonian}. More details about the thermal state preparation and its accuracy in the quantum simulation can be found in App.~\ref{Ising preparation}.
}
	\label{fig:Thermalstate}
\end{figure*}
 
As already pointed out in the introduction, a quantum simulation of Jarzynski's equality in the many-body regime faces some restrictive challenges.

First and foremost, the work distribution requires the measurement of initial and final eigenenergies~\cite{Talkner2007Fluctuation}, cf.~Eq.~\eqref{eq:work}. At the moment, this is only feasible for a suitable choice of the initial and final Hamiltonians,
and requires the ability to measure in their respective eigenbases.

We emphasize that the ability to apply general unitary transformations is a distinctive feature of digital quantum computers. In contrast, this is currently not possible, in general, for analog quantum simulators, such as cold atom systems~\cite{Bloch2008Many}; at present, this renders obtaining quantum work distributions in the many-body regime elusive on such platforms.
    
Second, the verification of Eq.~\eqref{eq:Jarzynski} requires the preparation of an initial thermal ensemble. Most approaches for Gibbs state preparation require an overhead of ancilla qubits~\cite{Wu2019Variational,Poulin2009Sampling,White2009Minimally,Zhu2020Generation,Riera2012Thermalization}. This uses up valuable qubits and makes the study of thermal states in the many-body regime difficult for the current generation of noisy intermediate-scale quantum (NISQ) devices.

Finally the number of required measurements increases with system size: According to~\cite{Jarzynski2006Rare,Halpern2016Number}, the number of shots $s$ scales approximately at least as
\begin{align}
    s\approx \mathrm{e}^{-\beta \braket{W_\mathrm{diss}}_{P(W)}}.
\end{align}
To get an estimate for $\braket{W_\mathrm{diss}}_{P(W)}$, we can use the fluctuation-dissipation relation~\cite{d2016quantum} to obtain the realation 
\begin{align}
 \braket{W_\mathrm{diss}}_{P(W)}\approx \frac{\beta}{2} \braket{W^2_\mathrm{diss}}_{P(W)}.
\end{align}
Recall that the variance of generic non-adiabatic work distribution scales linearly with the system size $L$~\cite{rigol2008thermalization}.
As a consequence, the number of required shots scales at least exponentially in the system size:
\begin{align}
    s= \mathcal{O}(\mathrm{e}^{-\beta^2 L})
\end{align}
We stress that this scaling can even become worse in the case of generic nonequilibrium protocols which can create long tails in the work distribution~\cite{Jarzynski2006Rare,Halpern2016Number}.

In the following, we describe in detail how we tackle each of these challenges. In Sec.~\ref{subsec:init_final_H}, we motivate our choice for the initial and final Hamiltonians. Section~\ref{subsec:jarzynski_protocol} elucidates our choice of nonequilibrium protocol, which is summarized diagrammatically in Fig.~\ref{fig:Thermalstate}.

\subsection{\label{subsec:init_final_H}Choice of initial and final Hamiltonian}

Jarzynski's equality depends neither on the specific form of the initial or final Hamiltonians, nor on the protocol we evolve the state with. However, a suitable choice of these ingredients can help to address the above-mentioned challenges, and enables its verification through a simulation on a quantum device operated in the many-body regime. 

\subsubsection{\label{subsec:work}Initial and final Hamiltonian}

Notice that determining the exponentiated work distribution requires measuring the initial and final energy, i.e., two-point measurements~\cite{Talkner2007Fluctuation}. In general, this can be achieved by applying a unitary transform to switchfrom the computational (i.e., the Pauli $z$-basis) into the energy eigenbasis, which is equivalent to diagonalizing the initial and final Hamiltonian, respectively. For generic systems, this requires a circuit of at least polynomial depth in the number of qubits $L$, using a quantum phase estimation algorithm~\cite{Abrams1999Quantum}. However, for systems equivalent to free fermionic models, circuits increasing logarithmically with system size suffice~\cite{CerveraLierta2018exactisingmodel}.

Apart from this practical restriction, the validity of Eq.~\eqref{eq:Jarzynski} imposes no further restrictions on the choice of initial and final Hamiltonians. 
Thus, we choose the initial Hamiltonian to be the transverse field Ising model with periodic boundary conditions~\footnote{For practical reasons, one has to add another Pauli string $\sigma^y_1 \sigma^z_2 \dots \sigma^z_{n-1} \sigma^y_n$. This term becomes negligible in the limit of large system sizes. Further details are discussed in App.~\ref{Ising preparation}.},
\begin{equation}\label{eq:initial Hamiltoniam}
    H_\mathrm{i}=\sum\nolimits_{\ell=1}^{L}\sigma^x_{\ell}\sigma^x_{\ell+1}+\sum\nolimits_{\ell=1}^L\sigma^z_\ell.
\end{equation}
This system is integrable and can be mapped to free fermions through a Jordan-Wigner transformation. It can be diagonalized using a shallow circuit as described in Ref.~\onlinecite{CerveraLierta2018exactisingmodel}. 

Although $H_\mathrm{i}$ is equivalent to a free-fermion model, its many-particle eigenstates are entangled and, therefore, feature genuine quantum correlations; thus, the circuit dynamics in the many-body regime goes beyond existing work on single- and few-particle models previously analyzed in the context of verifying Jarzynski's equality~\cite{an2015experimental,Bartalhao2014Experimental,smith2018verification,Gomez2020Experimental,Solfanelli2021Experimental,bassman2021computing}. 

As a final Hamiltonian, we chose the simple Hamiltonian
\begin{equation}\label{Final Hamiltonian}
    H_\mathrm{f}=\sum\nolimits_{\ell=1}^L\sigma^z_\ell,
\end{equation}
which is already diagonal in the computational basis,  such that measurements in the eigenbasis of $H_\mathrm{f}$ are straightforward. 
Overall, the choice of the initial and final Hamiltonians, which are either already diagonal in the computational basis, or where a circuit is known to diagonalize them in practice, enables us to perform a two-point measurement protocol to determine the work; this is essential to measure the left-hand side of the Jarzynski equality, cf.~Eq.~\eqref{eq:Jarzynski}. 

The partition functions for the initial and final ensemble (and thus the free energy difference) can be computed exactly for our choice of Hamiltonians.
The transverse field Ising model is equivalent to a system of non-interacting fermions
\begin{equation}
    H=\sum\nolimits_{k=-L/2+1}^{L/2} \omega_k a^\dagger_k a_k +E_c,
\end{equation}
with the energies $\omega_k$ defined in App.~\ref{Ising preparation}; the energy offset is given by $E_\mathrm{c}=1-\sqrt{2}$ (As mentioned above, we we work in units $k_\mathrm{B}=1$, $\hbar=1$).
The free energy therefore reads as
\begin{equation}
    F_\mathrm{i}=E_\mathrm{c}-\frac{1}{\beta}\sum\nolimits_{k=-L/2+1}^{L/2} \ln\bigl(1+\mathrm{e}^{-\beta \omega_k}\bigr).
\end{equation}
Since the energy offset $E_c$ enters as a constant additive term both in the definition of the work $W$ [Eq.~\eqref{eq:work}] and in the free energy $F_\text{i}$ above, and because the latter appear in the exponents on both sides of Eq.~\eqref{eq:Jarzynski}, we can ignore $E_\mathrm{c}$ in the following discussion.
The free energy of the final Hamiltonian is then
\begin{equation}
    F_\mathrm{f}=-\frac{L}{\beta}\ln\left[2\cosh\left(\frac{\beta}{2}\right)\right].
\end{equation}
In order to validate Eq.~\eqref{eq:Jarzynski}, it is therefore sufficient to focus on measuring the average exponentiated work.

\subsubsection{\label{subsec:Gibbs}Gibbs ensemble preparation: midcircuit measurements vs.~classical presampling}

To test the Jarzynski equality, we need to prepare a thermal ensemble w.r.t.~the initial Hamiltonian on the quantum simulator.
This is a difficult task on quantum devices in general since energy eigenstates are not necessarily accessible; however, the mapping of the Ising model to a model of non-interacting fermions allows us to use the initial energy measurement to prepare the thermal ensemble using a shallow circuit, as we now explain [see also parts (A) and (B) of the circuit in Fig.~\ref{fig:Thermalstate}]. 

The thermal density matrix for a system of non-interacting fermions can be written as 
\begin{eqnarray}
    \label{eq:gibbs_mb}
    \rho &=& \frac{1}{Z_\mathrm{i}}\sum_{m}\e^{- \beta E^\mathrm{i}_m} \ket{m}\bra{m} \\
    \label{eq:gibbs}
    &=&\frac{1}{Z_\mathrm{i}} \bigotimes_k \left(\e^{-\beta \omega_k} \ket{1}\bra{1}_k +\ket{0}\bra{0}_k\right),
\end{eqnarray}
where $\omega_k$ denotes again the energy of the single-particle eigenmodes.
In the case of free-fermions, each energy eigenstate $\ket{m}$ is uniquely defined by the collection of excited single-particle eigenmodes; This allows us to introduce an equivalence between bitstrings and eigenstates:
We can map an occupied single-particle eigenmode $k$ to the $k$-th qubit in the excited state, written as $\ket{1}_k$.
In the initial state, $\ket{0}^{\otimes L}$, all qubits are in the $\ket{0}$-state; thus, an excitation of the $k$-th mode is equivalent to applying an $X$-gate to qubit $k$.

\textbf{Midcircuit measurements:} starting from the product state $\ket{0}^{\otimes L}$, we can implement the Gibbs ensemble in two steps: 
(i) to prepare a thermal ensemble in the free fermion basis, we start from the state $\ket{0}$ and first apply a rotation gate $U_k$ to each qubit, of the form
\begin{equation}\label{rotation}
    U_k\equiv U\bigl(\theta_k\bigr)=\begin{pmatrix}    \cos\theta_k & -\sin\theta_k \\
	\sin\theta_k & \cos\theta_k \\\end{pmatrix},
\end{equation}
with $\theta_k$ implicitly defined by
\begin{equation}\label{rotationequation}
    \sin^2\theta_k=\frac{\e^{-\beta \omega_k}}{1+\e^{-\beta \omega_k}};
\end{equation}
(ii) we then perform a subsequent measurement of all qubits. The angles $\theta_k$ are chosen such that this projective measurement collapses the state with probability equal to that of the Gibbs ensemble; hence, the measurement statistics correspond to sampling from the Gibbs state in Eq.~\eqref{eq:gibbs}.
These two steps constitute part (A) of our circuit protocol, see Fig.~\ref{fig:Thermalstate}.

To complete the procedure, we have to apply a transformation from the energy to the computational eigenbasis.
This transformation consists of a fermionic Bogoliubov transform and a fermionic Fourier transform [cf.~App.~\ref{Ising preparation}]; these constitute part (B) of our circuit, see Fig.~\ref{fig:Thermalstate}. In the case of $L=2^k$ spins ($k\in\mathbb{N}$), the Fourier transform can be decomposed into fermionic SWAP gates and two-body Fourier transform gates, which can be efficiently implemented on a quantum computer~\cite{CerveraLierta2018exactisingmodel}.


We emphasize that, besides its use for Gibbs state preparation, the mid-circuit measurement simultaneously serves as a measurement of  $E^\mathrm{i}_m$, which is necessary to determine the work distribution.

\textbf{Classical presampling:} unfortunately, mid-circuit measurements are currently not feasible on all present-day quantum computing platforms. Wherever they are not available, we use classical presampling~\cite{CerveraLierta2018exactisingmodel,buffoni2020thermodynamics,hernandez2021experimental}. 
To this end, we prepare a randomly chosen many-body eigenstate $\ket{m}$ with probability equal to its Boltzmann weight, following Eq.~\eqref{eq:gibbs_mb}.
This is equivalent to sampling a bitstring directly from the Boltzmann distribution provided bitstring preparation can be performed with unit fidelity.

Since, at the end of the day, we want to perform a quantum simulation end-to-end, we avoid classical presampling whenever possible, and stick to midcircuit measurements for the largest system sizes $L=16$. Furthermore, classical presampling requires preparing different circuits corresponding to each different initial bitstring, while the use of mid-circuit measurements requires only one circuit for all possible initial bitstrings. As a consequence, this results in a speed-up for the execution of the simulation.

After having discussed the difficulties related to measuring work and thermal state preparation, we now move on to address the remaining challenges concerning the exponential scaling of measurement shots and free energy differences.

\subsection{\label{subsec:jarzynski_protocol}Nonequilibrium protocol}

In contrast to previous experiments, the local control over qubit interactions offers significant freedom in the choice of nonequilibrium protocol, cf.~part (C) of our circuit in Fig.~\ref{fig:Thermalstate}. 
Our goal is to devise a protocol that makes use of the intrinsic features of quantum computers in order to address the exponential scaling of the number of required measurements with the system size.

Because of the local nature of gates on digital quantum computers, we choose a circuit whose dynamics does not describe a parametric deformation between the initial and final Hamiltonians. This allows us to explore nonequilibrium protocols distinct from previous experiments~\cite{an2015experimental,Bartalhao2014Experimental}.

\begin{figure*}[t!]
	\centering
	\includegraphics{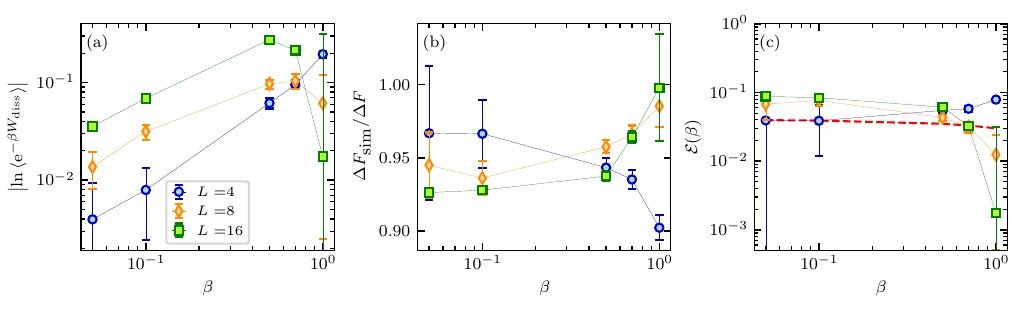}
	\caption{ 
	Testing Jarzynski's equality on a digital quantum computer using midcircuit measurements as a function of inverse temperature $\beta$, for three different systems sizes $L=4,8,16$. 
	\textbf{(a)} $\langle\e^{-\beta W_{\text{diss}}} \rangle_{P(W)}$ as a function of inverse temperature $\beta$. The unnormalized data depend on temperature and sytem size: deviations from Jarzynski's equality for a dissipation-free system are sensitive to the system size $L$ and grow with increasing inverse temperature.  
	\textbf{(b)} $\Delta F_\text{sim}/\Delta F$ as a function of inverse temperature $\beta$. The normalization largely removes the dependence  on temperature and sytem size.
	\textbf{(c)} 
	The same data, but now normalized by the maximum possible deviation [see Eq.~\eqref{Eq:normalized}]. The dashed red line shows the prediction from our theory derived in Sec.~\ref{sec:theory_analysis}.	Jarzynski's equality holds better than one part in ten, irrespective of the system size; the deviation agrees well with our theoretical prediction.
	We took $2^{16}$ measurements on ibmq\_guadalupe; the number of blocks in the circuit [cf.~Fig.~\ref{fig:Thermalstate}] is seven for $L=4, 8$, and three for $L=16$. Further technical details of the device can be found in App.~\ref{sec:Technical}.  
	}
	\label{fig:L16}
\end{figure*}

Instead, in our digital quantum simulations, we apply a protocol made out of $k$  sequential blocks, as shown in Fig.~\ref{fig:Thermalstate}(C). Each block consists of a layer of single-qubit Haar random unitary gates, followed by a sequential layer of CNOT-gates. In comparison to pure Haar random circuits, our circuit allows for a more native and shallow implementation on NISQ devices; moreover, $k=L-1$ blocks are sufficient to build up bipartite von Neumann entanglement in the system close to the Page value~\cite{Page1993Average}, as we demonstrate in a classical emulation in App.~\ref{sec:Entanglement}.

Our choice of a non-equilibrium protocol results in a work distribution $P(W)$, that approaches a Gaussian for system sizes $L\gtrsim 8$~[cf.~App.~\ref{sec:Entanglement}].
Thus, the tails of the work distribution exhibit a Gaussian decay and do not dominate the average of the exponentiated work~\cite{Jarzynski2006Rare,Halpern2016Number}.

However, the number of measurements to determine the distribution of the exponentiated work still scales exponentially with the system size $L$ and thus eventually gives rise to a bottleneck in testing Eq.~\eqref{eq:Jarzynski}. In our case, we find that for the case of $L=16$, $2^{16}$ measurements are sufficient. As we show below, this is feasible on current devices.

In order to determine the final energy (and with it, the work), we perform up to $2^{24}$ measurements on superconducting quantum computing architectures, and between $2^{12}$ (Quantinuum) and $2^{16}$ (IonQ) measurements on trapped ion platforms. The reason for the smaller number of measurements on trapped ion devices is the higher simulation cost of the quantum simulation due to lower protocol repetition rates, which restricts us to use less data points. For further details, see App.~\ref{sec:Comparison}.

Last, it is curious to note that the separation of parts (B) and (C) in our protocol [Fig.~\ref{fig:Thermalstate}] is somewhat arbitrary. On the one hand, part (B) belongs naturally together with part (A) in a thermal state preparation subprotocol. On the other hand, we can interpret part (B) as part of the nonequilibrium protocol (C) applied to the system. Since Jarzynski's equality holds for arbitrary protocols, equilibrium or nonequilibrium, the accuracy of implementation of part (B) is not crucial for the accuracy to which we verify Jarzynski's equality, provided the errors are systematic and are, thus, identical across trials. Note that this does not compromise the many-body character of our circuit: indeed, 
the entanglement created by the circuit reaches the maximal Page curve, cf.~Fig.~\ref{fig:operatorentanglement} in App.~\ref{sec:Entanglement}.

\section{Jarzynski equality and Crooks relation on a noisy quantum device}\label{sec:JarzynskiManybody}

After having discussed the details of the implementation of a test for the many-body quantum Jarzynski equality on quantum devices, we now provide simulation results for the test of Jarzynski's equality and Crooks' relation on present-day quantum computing platforms.

\subsection{Jarzynski relation in the few- and many-body regime}

The results presented below are obtained in experiments on the ibmq\_guadalupe superconducting quantum processor~\cite{IBM}. A detailed comparison with other digital quantum architectures is included in App.~\ref{sec:comparison}. For the technical details of the various devices including gate fidelities and relaxation times $T_1$ and $T_2$, we refer the interested reader to App.~\ref{sec:Technical}.

In Fig.~\ref{fig:L16} we display the deviations from Jarzynski's equality as a function of the inverse temperature. The three panels show the same data, but viewed through the lens of different quantifiers of the deviation.

Let us open up the discussion in Fig.~\ref{fig:L16}~(a) by introducing the plain deviation quantifier $\ln\bigl\langle\e^{-\beta W_{\text{diss}}}\bigr\rangle_{P(W)}$, which vanishes whenever Jarzynski's equality holds.
The curves show an approximately linear scaling with the inverse temperature $\beta$ in the high-temperature limit. The dominant contributions to this deviation quantifier originate from violations of double-stochasticity with equal contributions from all final energy eigenstates. 
By contrast, for inverse temperatures $\beta \gtrsim 1$, the deviations are determined by processes involving the ground state of the final system. As a consequence, the deviations for large $\beta \gtrsim 1$ will converge to a constant value, which depends on the concrete effect of energy dissipation on the process. For a detailed discussion, see App.~\ref{sec: high- and low temperature expansion}. 
The comparison between the three system sizes in Fig.~\ref{fig:L16}~(a) reveals increasing absolute deviations from Jarzynski's equality with increasing~$L$. We suspect that this is due to the increasing number of qubits to be read out: a primary contribution to the overall error arises from measurement errors, and hence the error in the work measurement scales linearly with system size. We provide a more detailed analysis of various errors in Sec.~\ref{sec:theory_analysis} and App.~\ref{sec:Error}. For a meaningful and systematic comparison of the results for different system sizes we, therefore, consider different quantifiers for relative deviations in panels (b) and (c).

Figure~\ref{fig:L16}~(b) shows the ratio ${\Delta F_\text{sim}}/{\Delta F}$, introduced in previous work~\cite{an2015experimental,Solfanelli2021Experimental}. Here, 
\begin{align}
    \Delta F_\text{sim}=-\frac{1}{\beta}\ln\bigl\langle{\e^{-\beta W}}\bigr\rangle_{P(W)}\ . 
\end{align}
is the free energy difference obtained from the distribution average of the exponentiated work and $\Delta F$ is the theoretical prediction for the free energy difference.
Any deviations from the ideal, dissipation-free case, are indicated by deviations of the ratio from unity. We observe ${\Delta F_\text{sim}}/{\Delta F}>90\%$ for all inverse temperatures and only a weak system-size dependence, since the latter is absorbed by the scaling of the denominator.

Since $\Delta F$ is a protocol-dependent quantity and not immediately accessible for a given setting, we introduce another normalization, which facilitates a straightforward quantitative comparison with previous experiments and which is tailored to quantify the amount of energy dissipation in the system.
For this purpose, we consider the process of dissipative decay to the ground state of the final system as the natural reference for deviations from the Jarzynski equality, because it constitutes its worst possible violation. The resulting work obtained by the two-point measurement scheme when starting from an initial energy $E^{\text{i}}$ is $W=E^\text{f}_0-E^\text{i}$.
Accordingly, we introduce
\begin{equation}
\label{Eq:idle process}
    { \mathrm{e}^{-\beta W_\mathrm{decay}}}\equiv\overline{\mathrm{e}^{-\beta\bigl(E^\mathrm{f}_0-E^\mathrm{i}\bigr)+ \beta \Delta F_\mathrm{sim}}}\ ,
\end{equation}
where the right-hand side is the ratio of the exponentiated energy differences for the purely dissipative process, and $\Delta F_\mathrm{sim}$ for the true process simulated on the quantum computer. The bar $\overline{(\cdot)}$ denotes an average over the initial thermal ensemble.

This allows us to define the relative deviation from the theoretical prediction,
\begin{equation}
\label{Eq:normalized}
    \mathcal{E}\bigl(\beta\bigr)=\left |\frac{\ln{\bigl\langle\e^{- \beta W_{\text{diss}}}\bigr\rangle_{P(W)}}}{\ln \mathrm{e}^{-\beta W_\mathrm{decay}}}\right |.
\end{equation}
This quantity is bounded from below and above. Whenever Jarzynski's equality holds, we have $\mathcal{E}(\beta)=0$. On the other hand, for a purely dissipative process, $\mathcal{E}(\beta)=1$.
In contrast to previous deviation quantifiers, the normalization in $\mathcal{E}(\beta)$ by the worst-case scenario allows us to directly compare the amount of dissipation in our simulations and previous experiments, cf.~Table~\ref{Expcomparison} in the Discussion. Moreover, small deviations from the equality can be resolved logarithmically, while the upper bound simultaneously justifies a quantitative interpretation.

\begin{figure}[t!]
	\centering
	\includegraphics[width=1\columnwidth]{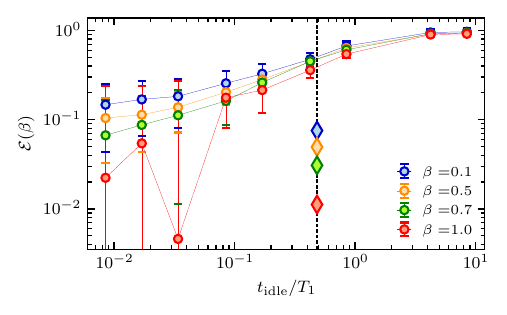}
	\caption{Relative deviation from Jarzynski's equality as a function of the waiting time $t_{\text{idle}}$~(solid lines), for four different inverse temperatures~$\beta$, at $L=8$. The $x$-axis is normalized by the average waiting time $T_1= 118$~$\mu$s~\cite{Notefig4}. For comparison, the diamonds denote the results for our protocol shown in Fig.~\ref{fig:L16}(b), placed at their respective execution time [dashed vertical line]. The application of gates between the two measurements in the protocol [cf.~Fig.~\ref{fig:Thermalstate}] improves the accuracy of our results by up to more than one order of magnitude.
	}
	\label{fig:Randomlayers}
 
\end{figure}

The relative deviation from Jarzynski's equality as defined in Eq.~\eqref{Eq:normalized} is plotted in Fig.~\ref{fig:L16}~(c). Consistently with Fig.~\ref{fig:L16}~(b) the results are largely independent of the inverse temperature $\beta$ and the relative deviations do not exceed $10\%$.
As for the previous normalization ${\Delta F_\text{sim}}/{\Delta F}$, $\mathcal{E}(\beta)$ exhibits only a weak system-size dependence.
Because increasing the system size requires a larger circuit depth via the sub-circuits (B) and (C) in Fig.~\ref{fig:Thermalstate}, one could na\"ively expect that scaling to large system sizes is quickly hampered by dissipation effects which accumulate with increasing circuit size. However, this appears not to be the case in the observed behavior in Fig.~\ref{fig:L16}, where we show data up to $L=16$ qubits. Therefore, we now briefly investigate our circuit's susceptibility to energy dissipation.

The purely dissipative process leading to the right-hand-side of Eq.~\eqref{Eq:idle process} can be emulated on a NISQ device by applying a so-called "idle process" \cite{sommer_many-body_2021}: this is a similar process as the one described in Fig.~\ref{fig:Thermalstate}, but with the circuit parts (B) and (C) replaced by free evolution for a variable duration $t_\mathrm{idle}$. In other words, the two measurements at the end of circuit part (A) and in part (D) are separated by the idle time $t_\mathrm{idle}$.
Hence, to access the regime of validity of Eq.~\eqref{Eq:idle process}, we have to apply an idle process of time $t_\mathrm{idle} \gg T_1$, with $T_1$ being the dissipation time. 

Figure~\ref{fig:Randomlayers} shows the relative deviation from Jarzynski's equality for an idle process, as a function of the waiting time $t_\mathrm{idle}$ for $L=8$ qubits. As expected, for large waiting times $t_\mathrm{idle}\gtrsim T_1$, $\mathcal{E}\to 1$ reaches the limit of a pure dissipative process.
We now want to compare the time required to reach the purely dissipative regime with the execution time of the circuit from Fig.~\ref{fig:Thermalstate}.
Before we do this, notice first that the average dissipation time of ibmq\_guadalupe is $T_1 \approx 118$~$\mu$s, while the execution time for our circuit~[Fig.~\ref{fig:Thermalstate}, (A) to (D)] is $T_c=57$~$\mu$s for $L=8$ qubits; hence ${T_c}/{T_1} \approx 0.48$ [cf.~ Diamonds on dashed vertical line in Fig.~\ref{fig:Randomlayers}] [for comparison, for 16 qubits we have ${T_c}/{T_1}\approx 2.95$]. Furthermore, the normalized deviation of the quantum Jarzynski equality Eq.~\eqref{Eq:normalized} does not depend on the number of entangling blocks, as is shown in Fig.~\ref{fig:Blocksize} in App.~\ref{subsec:blockdependence}.

Comparing the deviation values with the idle process, the accuracy of the process from Fig.~\ref{fig:L16} appears quite striking, since the deviation there is up to more than one order of magnitude smaller than an idle protocol of the same waiting time, $t_\mathrm{idle}=T_c$.
In fact, the deviation values are comparable to those of the idle process with the smallest idle time investigated, from which we can deduce that the main source of deviations from Eq.~\eqref{eq:Jarzynski} originates in part (D) of the protocol.

To sum up, our nonequilibrium circuit has the property of preventing energy dissipation effects from accumulating, despite increasing circuit execution time; this is, in turn, reflected in Jarzynski's equality being obeyed to a high degree of accuracy even in the many-body regime of $L=16$ qubits. We will discuss this observation in more detail in Sec.~\ref{sec:theory_analysis}.

\begin{figure}[t!]
	\centering
	\includegraphics[width=1\columnwidth]{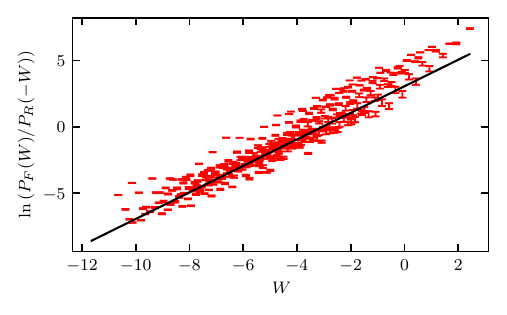}
	\caption{Test of the Crooks relation $\ln\left(\frac{P_{F}(W)}{P_{R}(-W)}\right)$ as a function of the work $W$. 
    The black line denotes the theoretical prediction for a noiseless device, Eq.~\eqref{Crooks element log}. Although the results from the quantum simulation follow the theory prediction, deviations indicate the violation of double-stochasticity.
    The data shown here are for $\beta=1.0$ , $2^{24}\sim 1.7 \times 10^7$ shots and $L=8$ qubits on ibmq\_guadalupe. 
	}
	\label{fig:Crooks8}
\end{figure}

\subsection{Crooks relation in the many-body regime}

While Jarzynski's equality is a statement about work-averaged processes, we can also check to what extent its infinitesimal [or in-sample] version, the Crooks relation~\cite{Crooks1999Entropy,tasaki2000jarzynski}
\begin{equation}
\label{Crooks element log}
    \ln\left(\frac{P_F(W)}{P_R(-W)}\right)=\beta (W-\Delta F)\ ,
\end{equation}
holds [see App.~\ref{sec:theory} for a derivation]. To the best of our knowledge, the latter was only tested in two-level quantum systems so far~\cite{Bartalhao2014Experimental,Batalhao2015Irreversibility}.

For a process described by a sequence of unitary gates, and ignoring noise for the time being, the backward process can be implemented in a straightforward way by reversing the gate order and taking the inverse of each individual gate.

We tested the Crooks relation for $L=8$ qubits, using $2^{24}\approx 1.7 \times 10^7$ measurements to reduce statistical fluctuations due to a limited number of measurements.
The corresponding data are shown in Fig.~\ref{fig:Crooks8}. The validity of the Crooks relation requires that all data points lie on the black line, apart from statistical fluctuations arising from a finite number of measurements. Indeed, the data follow the trend predicted by the Crooks relation.
However, the deviation between the theory prediction and the measurement outcomes cannot solely be explained by the statistical uncertainty due to a limited number of measurements:
The ratio between occurrences of forward and backward processes in Eq.~\eqref{Crooks element} is up to one order of magnitude larger than expected due to statistical errors. These deviations are a consequence of the presence of energy dissipation in the device.

\section{\label{sec:theory_analysis}Theoretical analysis}

In order to understand the small deviations in the validation measurement for the Jarzynski equality and the strong fluctuations in the Crooks relation, we analyze a simplified single-qubit toy model subject to a periodic application of a single-qubit rotation gate, instead of a random sequence of gates. As we discuss below, the main results of this analysis extrapolate also to the protocol from Fig.~\ref{fig:Thermalstate}.

\subsection{Single-qubit case}

Consider first a single qubit. The Hilbert space is two-dimensional, such that the state can be expressed by a vector $\ket{\psi}=(u_1,u_2)$. To be more concrete, $\ket{\psi_G}=(1,0)$ defines the "physical" ground state of the qubit, whereas $\ket{\psi_E}=(0,1)$ denotes an excited state.

Consider the  single-qubit density matrix $\rho$:
\begin{equation}
    \rho= \begin{pmatrix}
    \rho_{11}&\rho_{12} \\\rho_{21} &\rho_{22} 
    \end{pmatrix}.
\end{equation}
We analyze the repeated [periodic] application to the qubit of a unitary gate $U$ of the form
\begin{equation}
    U=   \begin{pmatrix}
    \cos\alpha&-\sin\alpha \\\sin\alpha &\cos\alpha 
    \end{pmatrix}.
\end{equation}
The density matrix vector evolves according to 
\begin{equation}
    \rho \rightarrow U \rho U^\dagger.
\end{equation}
Let us assume that the excited state has a finite lifetime $T_1$; thus, we can use a single amplitude damping channel~\cite{nielsen2002quantum} to describe this effect. The time evolution over one period is then given by
\begin{equation}
  \mathcal{F}\bigl(\rho\bigr)=\sum\nolimits_{i=1}^2\mathcal{K}_i \bigl(U \rho U^\dagger \bigr) \mathcal{K}_i^\dagger.
\end{equation}
 
Here $\mathcal{F}\bigl(\rho\bigr)$ is a completely positive trace-preserving map on the space of density matrices with a unique largest eigenvalue $\lambda=1$ for $p>0$. 
The Kraus operators $\mathcal{K}_i$ describing the damping process are given by
\begin{equation}
  \mathcal{K}_1= \begin{pmatrix}
    1 &0 \\ 0&\sqrt{1-p} 
    \end{pmatrix},\quad\mathrm{and}\quad\mathcal{K}_2= \begin{pmatrix}
    0 &\sqrt{p} \\ 0&0 
    \end{pmatrix}.
\end{equation}
The parameter $p$ is the fraction of the excited state population which decays to the ground state after one application of $\mathcal{F}$. It is related to the relaxation time $T_1$ and the gate time $T_g$ via
\begin{equation}
\label{eq:defp}
    p=1- \exp\left(- \frac{T_g}{T_1}\right).
\end{equation}
Repeating the process $N$ times, the density matrix evolves in the long time limit as
\begin{equation}\label{eq:process}
  \lim_{N\rightarrow \infty} \mathcal{F}^N\bigl(\rho\bigr)= \rho_0.
\end{equation}
A simple calculation gives, for $p>0$,
\begin{equation}
  \rho_0=\begin{pmatrix}
    1/2+f(p,\alpha) &g(p,\alpha)\\
    g(p,\alpha)& 1/2-f(p,\alpha)\\
    \end{pmatrix}, 
\end{equation}
with 
\begin{equation}\label{eq:steady state correction}
    f(p,\alpha)=\frac{p\left[1-\sqrt{1-p}\cos(2\alpha)\right]}{2\Bigl[1-p+\sqrt{1-p}\Bigr]\Bigl[1-\cos(2 \alpha)\Bigr]},
\end{equation}
and
\begin{equation}
    g(p,\alpha)=\frac{p\sqrt{1-p}\sin(2\alpha)}{2\Bigl[1-p+\sqrt{1-p}\Bigr]\Bigl[1-\cos(2 \alpha)\Bigr]}.
\end{equation}
Jarzynski's equality holds for $f(p,\alpha)=0$, which is indeed the case at $p=0$.

Our quantum simulations testing Jarzynski's equality are in the regime $T_g/T_1 \ll 1$. In order to understand this regime, we perform an expansion of $f(p,\alpha)$ around $p=0$; together with Eq.~\eqref{eq:defp} this gives
\begin{equation}
\label{eq:limit}
    f(p,\alpha)=\frac{1}{4}p +\mathcal{O}\bigl(p^2\bigr)=\frac{1}{4} \frac{T_g}{T_1} +\mathcal{O}\left(\frac{T_g^2}{T_1^2}\right).
\end{equation}
This regime is approached in the case of an infinitely fast drive, as can be seen from the relation in Eq.~\eqref{eq:limit}. Note that the actual time of the protocol does not matter in this case, since this analysis holds in the limit of infinitely many periods $N\rightarrow \infty$. Indeed, there is no obstruction if the protocol time greatly exceeds the relaxation time scale $T_1$.

\subsection{Extension to multiple qubits}

Although we restricted the analysis to the case of a single qubit with a periodic protocol, we can apply it to random gates and multi-qubit systems.
For simplicity, let us only take into account the independent decay of qubits in the excited state to their ground state, and ignore many-particle effects for a moment. Although this is a crude simplification, it will turn out that it gives quantitative good results, such that this approximation is well justified a posteriori.
The relevant physical parameter is $p$, as defined in Eq.~\eqref{eq:defp}. The application of single-qubit gates can be interpreted as a drive applied to the system, which repopulates excited states and thus compensates for the deviations from Jarzynski's equality caused by the energy dissipation process.

\begin{table*}
\label{Expcomparison}
\centering
\begin{tabular}{||c| c |c |c |c||} 
 \hline
 \textbf{Experiment} & \textbf{Experimental platform} & \textbf{System size} $L$ & \textbf{Inverse temperature} $\beta$ & \textbf{Relative deviation} $\mathcal{E}(\beta)$ \\ [0.5ex] 
 \hline\hline
 \rowcolor{Gray}
 This work  & superconducting qubits & 16  &$0.7$&$0.061(3)$ \\ 
 \hline
 Ref.~\onlinecite{bassman2021computing}& superconducting qubits&3 &$1.0$&  $0.03(1)$\\
   \hline
 Ref.~\onlinecite{Solfanelli2021Experimental}& superconducting qubits&1 & $1.0$& $0.02(2)$  \\
 \hline
   Ref.~\onlinecite{an2015experimental}&trapped ions: vibrational modes &1 &1.13& 0.02(2) \\
\hline
  Ref.~\onlinecite{smith2018verification}&trapped ions: two hyperfine levels &1 &1.3& 0.03(9) \\
\hline
  Ref.~\onlinecite{Bartalhao2014Experimental}& NMR &1 &$0.15$ &0.17(8) \\
 \hline
   Ref.~\onlinecite{cerisola2017using}& Hyperfine levels of $^{87}$Rb &1 &1.75 &0.00(4) \\
 \hline
  Ref.~\onlinecite{hernandez2021experimental}& NV centers&1 & $0$ &0\\
  \hline \hline
   \end{tabular}
\caption{
Comparison of different experiments for the validation of the quantum Jarzynski equality, using the normalized deviation defined in Eq.~\eqref{Eq:normalized}. The relative accuracy $\mathcal{E}$ of our simulations is comparable with previous experiments on few-qubit systems. More details for the extraction of the data and the determination of $\beta$ are provided in the appendix~\ref{sec:comparison}.
}
\end{table*}

To make a quantitative estimate for our experiment, we determine the parameter $p$ and the measurement error of ibmq\_guadalupe.
The execution time of single-qubit gates is negligible in comparison to two-qubit gates $T_g$, so $p$ is given by:
\begin{equation}\label{eq:pGuad}
    p=1-\mathrm e^{-\frac{T_g}{T_1}}=3 \times 10^{-3},
\end{equation}
see App.~\ref{sec:Technical} for the characteristic time scales of the device.
Thus, in our approximation, the effect of 2-qubit entangling gates only enters via the gate time $T_g$. 
Furthermore, averaging over the angle $\alpha$ for each single rotation gate, $f(p,\alpha)$ reduces to 
\begin{equation}
f_0(p)=\frac{p}{2\Bigl[1-p+\sqrt{1-p}\Bigr]}\approx 8 \times 10^{-4}.
\end{equation}

In addition, we have to factor in the dissipation during the measurement process, i.e., excited qubits which are misidentified to be in the ground state. In the following, we neglect the opposite case, i.e., a qubit in the ground state misidentified to be in the excited state. 
For the device ibm\_guadalupe, we get an approximate measurement error of $p_0\approx 0.04$.
The steady-state probability  for each single qubit to be in the excited state is then given by
\begin{equation}
p_\text{exc}=\left(\frac{1}{2}-f_0(p)\right)\left(1-p_0\right)=\frac{1}{2}-p_0-f_0(p)-f_0(p)p_0.
\end{equation}

Since $f_0(p)\ll p_0$, the model predicts that the deviations from our experiment are almost exclusively due to measurement imperfections.
The analytically estimated deviation from a dissipation-free evolution is indicated as a red dashed line in Fig.~\ref{fig:L16}. Taking the simplifications of our analysis into account, the result agrees quantitatively well with our simulation results and suggests that measurement errors are indeed the dominant source for deviations from the quantum Jarzynski equality.

\subsection{Relation to DiVincenzo's third criterion}

The compensation for energy dissipation is reminiscent of DiVicenzo's criteria for quantum computing~\cite{DiVincenzo2000The}. To be more concrete, the third criterion states that scalable quantum computing requires long decoherence times in comparison to the time scale of operational gates. 

In our case, we are not interested in  scalable fault-tolerant quantum computing, but in a weaker question, namely under which conditions the quantum Jarzysnki equality Eq.~\eqref{eq:Jarzynski} is fulfilled. This only requires long \textit{dissipation} timescales. In contrast to fault-tolerant quantum error correction, it is not necessary to actively correct for errors; the dynamics itself already compensates for the energy dissipation. Put differently, the application of single-qubit gates can be interpreted as a re-population of excited states.
The quantum Jarzynski equality, thus, holds in an effective way, although every single component of the dynamics is spoiled by energy dissipation.

\section{Discussion \& Outlook}\label{sec:Discussion}

We proposed a protocol to test the quantum Jarzynski equality and the Crooks relation in the many-body regime on near-term quantum computing devices, in the presence of different errors. We pushed the state of the art for the quantum simulation of Jarzynski's equality up to 16 qubits, and for the Crooks relation -- to 8 qubits, respectively. We identified the implementation of two-point work measurements, the preparation of a thermal ensemble on a digital quantum simulator, and the exponential growth of the required number of measurements as the major practical challenges to reach the many-body regime.

To address the first two challenges, we developed a protocol sequence that prepares a canonical ensemble by using mid-circuit measurements. While for small system sizes classical presampling is still feasible, mid-circuit measurements allow us to prepare thermal ensembles for up to $L=16$ qubits without running multiple independent experiments and without any overhead incurred by using ancilla qubits. Taking the transverse field Ising model and an $S^z$-model as exemplary initial and final Hamiltonians, respectively, we performed energy measurements with circuits scaling at most logarithmically in the system size $L$. This approach is exact for Hamiltonians equivalent to single-particle systems; it is currently an open question whether, to what accuracy, and under which conditions, one could prepare thermal ensembles for more general Hamiltonians using similar protocols.
A possible route could be to use variation quantum eigensolvers for the preparation of thermal ensembles for non-integrable systems~\cite{selisko2022extending}.

While the exponential (in the system size) number of projective measurements still appears as the dominant bottleneck in testing the quantum Jarzynski equality when increasing the number of degrees of freedom, we were able to collect enough measurement data to test the latter for up to $L=16$ qubits.
As a side product, our protocol in Fig.~\ref{fig:Thermalstate} reveals the ingredients of Eq.~\eqref{eq:Jarzynski} in a simple way: the quantum Jarzynski equality is a relation between initial and final eigenenergies, and a double-stochastic transfer matrix connecting them. Any other terms, including the transformation to a computational basis, can be absorbed into the transition process itself.
The quantum Jarzynski equality thus probes only the double-stochasticity of the process and not the accuracy of different parts of the protocol.

Let us also compare our results to previous experiments on the quantum Jarzynski equality, cf.~Table~\ref{Expcomparison}. To this end, we first extract the results for the left-hand side of Eq.~\eqref{eq:Jarzynski}, and the theory prediction of the free energy of the corresponding experiments; then we compute the maximum deviation of an idle process, cf.~Eq.~\eqref{Eq:idle process}, which gives us the normalized deviation defined in Eq.~\eqref{Eq:normalized} for each experiment [cf.~Appendix~\ref{sec:comparison} for details]. The comparison shown in Table~\ref{Expcomparison} clearly suggests that the accuracy of our results is comparable with most of the earlier experiments; however, we reach an order of magnitude larger system sizes, where quantum many-body effects become pronounced. 

Moreover, in contrast to previous experiments, the protocol duration of our circuits is comparable to, or even exceeds, the average dissipation time $T_1$ of the NISQ devices; therefore, energy dissipation is no longer negligible. 
We checked that our results do not depend on the specific choice of randomness in our protocol. 
Furthermore, we demonstrated that the relative accuracy of our results is almost independent of system size and circuit depth.
We also developed a theoretical model which predicts the empirical observations in a quantitative manner.
By employing a fast drive that compensates for energy dissipation, we thus found the Jarzynski equality to be effectively valid in this regime, even though the dynamics is not doubly stochastic.

While these observations are interesting on their own, our work demonstrates two  promising practical applications:
First, testing the equality can be used to investigate errors on NISQ devices in new ways: Since Jarzynski's equality is only sensitive to processes violating double-stochasticity, it can be used to quantify and single out this effect. Even though simpler protocols to determine the dissipation time $T_1$ of devices exist~\cite{nielsen2002quantum}, our approach~can be used as the generalization of these approaches to the many-qubit regime: Jarzynski's equality can not only be used to detect the decay of excited states as shown in our analysis in Sec.~\ref{sec:theory_analysis}, but it is, more generally, sensitive to any violation of double-stochasticity. 
It remains an open question for future studies to investigate to what extent this allows us to refine our understanding of correlated error processes on modern quantum devices.
For instance, the commonly used protocol for measuring $T_1$ can be interpreted as special cases of testing the Jarzynski equality for a single spin system, as we show explicitly in App.~\ref{sec:T_1}. 
Second, our analysis emphasizes important limitations concerning the measurement of free energy differences -- one of the promising applications of the Jarzynski equality: in generic non-integrable systems, the exact work distributions can only be extracted by diagonalizing the circuits, which requires at least polynomially deep circuits using quantum phase estimation~\cite{Vogel1989Determination} and is thus infeasible in the many-body regime. 
It remains an exciting question for future research to find approximations to Eq.~\eqref{eq:Jarzynski}, which allow for a scalable method to extract free energies using the quantum Jarzynski equality in the many-body regime~\cite{bassman2021computing}.

\section{Acknowledgments}

We thank Florian Mintert, Adam Smith, and Nicole Yunger-Halpern for inspiring discussions. We would also like to thank the referees for their comments and useful advice on our manuscript.
This project was supported by the Deutsche Forschungsgemeinschaft (DFG) through the cluster of excellence ML4Q (EXC 2004, project-id 390534769). We acknowledge support from the QuantERA II Programme that has received funding from the European Union’s Horizon 2020 research innovation programme (GA 101017733), and from the Deutsche Forschungsgemeinschaft through the project DQUANT (project-id 499347025).
M.B.~was supported by the Marie Sk\l{}odowska-Curie grant agreement No 890711. 
To perform the quantum simulations: We acknowledge support from the Microsoft Azure Quantum Credit Program.
This work was supported by the AWS Cloud Credit for Research program.
We acknowledge the use of IBM Quantum services for this work; the views expressed are those of the authors, and do not reflect the official policy or position of IBM or the IBM Quantum team. 
This research used resources of the Oak Ridge Leadership Computing Facility, which is a DOE Office of Science User Facility supported under Contract DE-AC05-00OR22725.

\appendix

\section{\label{sec:theory}Quantum Jarzynski Equality and Crooks Relation}

Let us recall the derivation of the quantum Jarzynski equality~\cite{kurchan2001quantum,Mukamel2003Quantum}.
Consider a system described by a Hamiltonian $H_\mathrm{i}$ with eigenstates $H_\mathrm{i}|m^\mathrm{i}\rangle{=}E^\mathrm{i}_m|m^\mathrm{i}\rangle $, coupled to a thermal reservoir of inverse temperature $\beta$. The system is thus described by a thermal ensemble with the density matrix $\rho=Z_\mathrm{i}^{-1}\sum_{m}\e^{-\beta E^\mathrm{i}_m}\ket{m^\mathrm{i}}\bra{m^\mathrm{i}}$, and $Z_\mathrm{i}=\sum_{m} \e^{- \beta E^\mathrm{i}_m}$ is the partition function.
We now decouple the system from the reservoir, and let it evolve according to a dynamical process $U$~(not necessarily unitary, see below). At the end of this process, the instantaneous final Hamiltonian $H_\mathrm f$ of the system has eigenstates $H_\mathrm{f}|n^\mathrm{f}\rangle{=}E^\mathrm{f}_n|n^\mathrm{f}\rangle$. We denote by $K_{m\rightarrow n}$ the transition probability from the initial eigenstate $|m^\mathrm{i}\rangle $ to the final eigenstate $|n^\mathrm{f}\rangle$.

While the dynamical process $U$ need not be unitary, we require that the transition probabilities satisfy the following two sum rules:
\begin{equation}
    \label{eq:Sumrules}
    \sum\nolimits_{n} K_{m\rightarrow n}=1,\quad \forall m,~~~\mathrm{and}~~~\sum\nolimits_{m}  K_{m\rightarrow n}=1, \quad \forall n.
\end{equation}
The left-hand equality reflects the conservation of probability. The right-hand sum rule is less obvious and implies the so-called double-stochasticity of the matrix $K_{mn}$; this condition is fulfilled for unitary dynamics $K_{m\rightarrow n} = |\langle n^\mathrm{f} |U| m^\mathrm{i}\rangle |^2$, but is also conserved throughout evolution in the presence of additional decoherence noise~\cite{smith2018verification}. By contrast, energy dissipation violates the right-hand sum rule condition~\cite{smith2018verification,gardas2018quantum}.

Using these definitions, we can now prove the quantum Jarzynski equality for the process introduced above:
\begin{eqnarray}
    \Bigl\langle\e^{-\beta\Delta W}\Bigr\rangle_{P(W)}&=&\frac{1}{Z_\mathrm{i}}\sum_{m,n}\e^{-\beta E^\mathrm{i}_m} K_{m\rightarrow n} \e^{-\beta W_{nm}} \nonumber \\
    &=& \frac{1}{Z_\mathrm{i}} \sum_{m,n} K_{m\rightarrow n} \e^{-\beta E^\mathrm{f}_n}=\frac{1}{Z_\mathrm{i}}\sum_{n}\e^{-\beta{E^\mathrm{f}_n}} \nonumber \\ 
    &=& \frac{Z_\mathrm{f}}{Z_\mathrm{i}}=\e^{-\beta\Delta F},
\end{eqnarray}
where we used $W_{nm}{=}E^\mathrm{f}_n{-}E^\mathrm{i}_m$ according to Eq.~\eqref{eq:work} and the double-stochasticity of $K_{m\rightarrow n}$ in the second line, and the definition of free energy $F_\mathrm{i}=-\beta\ln Z_\mathrm{i}$ in the last line.
We note that Jarzynski's equality imposes no restrictions on the initial and final Hamiltonians; in particular, they need not be identical.

There exists an infinitesimal (i.e., work-resolved) version of Jarzynski's equality, called the Crooks relation~\cite{Crooks1999Entropy,tasaki2000jarzynski}. 
The Crooks relation states that
\begin{equation}
\label{Crooks element}
    \frac{P_F(W)}{P_R(-W)}=\mathrm{e}\,^{\beta (W-\Delta F)}.
\end{equation}
Here $P_F(W)$ denotes the probability of extracting an amount of work $W$ for a given (so-called forward) process and $P_R(-W)$ for the reverse (or backward) protocol, which can be expressed as 
\begin{eqnarray}
    P_F(W)&=&\frac{\e^{-\beta{E^\mathrm{i}_m}}}{Z_\mathrm{i}}K_{m\rightarrow n} \bigg|_{E^\mathrm{i}_m-E^\mathrm{f}_n=W}\;,\\
    P_R(-W)&=&\frac{\e^{-\beta{E^\mathrm{f}_n}}}{Z_\mathrm{f}}K_{m\leftarrow n} \bigg|_{E^\mathrm{f}_n-E^\mathrm{i}_m=-W}\;.
\end{eqnarray}
The Jarzynski equality follows by rearranging the Crooks relation, and integrating it over the work $W$.

\section{Comparison of different NISQ architectures}
\label{sec:Comparison}
In the following section, we give a brief overview of the different quantum computing platforms, the noise they are affected by, and their significance for the test of Jarzynski's equality.

\subsection{\label{subsec:archs}NISQ architecture characteristics}

In order to test the effect of different noise types, we run our circuits on five different devices using two different architectures: superconducting qubits (ibm\_perth, ibmq\_guadalupe and Rigetti Aspen-11), and trapped ion platforms~(Quantinuum H1 and an 11-qubit system of IonQ).  We extract the exponential of the work, Eq.~\eqref{eq:work}, to test the Jarzynski equality, cf.~Eq.~\eqref{eq:Jarzynski}. As mentioned above the latter is valid also in the presence of noise which does not violate double-stochasticity, and is only sensitive to errors that violate the second sum rule in Eq.~\eqref{eq:Sumrules}. The size of the deviation from the theoretical prediction Eq.~\eqref{eq:Jarzynski}, valid for an ideal dissipationless device, gives us therefore information about the amount processes violating double-stochasticity during the simulation.

The quality of qubits is often measured by means of the average gate times $T_g$ and the relaxation times $T_1$ and $T_2$.  The timescale of dephasing errors, $T_2$, is not relevant for our purposes, since depolarizing errors do not violate double-stochasticity~\cite{Mukamel2003Quantum,smith2018verification}. Only the thermal relaxation time $T_1$ or, more concretely, the ratio $q={T_g}/{T_1}$  matters, as it sets the decay rate for excited states, cf.~Sec.~\ref{sec:JarzynskiManybody}. While the two-qubit fidelities for all architectures fall between 95\% and 99.5\%, the $q$-factor depends strongly on the underlying architecture.

As discussed in Sec.~\ref{sec:theory_analysis}, the accuracy to which the Jarzynski equality~Eq.~\eqref{eq:Jarzynski} holds in experimental setups, depends on 
(a) the ratio between gate time $T_g$ and the relaxation time $T_1$, 
(b) measurement errors, and 
(c) statistical errors due to a finite number of measurements. The statistical error reduces with the square root of the number of measurements. Thus, by testing to what accuracy Eq.~\eqref{eq:Jarzynski} holds, we gain information about processes violating double-stochasticity in the quantum device. The two-qubit gate time for IonQ-devices is $T_g\sim 200$~$\mu$s, and the relaxation time $T_1\sim 10^7$~$\mu$s, resulting in a factor $q_{\text{IonQ}}~\sim \mathcal{O}(10^5)$. On the other hand, the timescale for  IBM-platforms are $T_g\sim 400$ ns, $T_1\sim 160$~$\mu$s, yielding $q_{\text{IBM}}~\sim 400$. Based on these estimates, we anticipate obtaining significantly smaller deviations from Jarzynski's equality~Eq.~\eqref{eq:Jarzynski} on trapped ion compared to superconducting platforms. Furthermore, due to the high-quality factor $q_{\text{IonQ}}$ we can assign deviations from the Jarzynski equality on trapped ion platforms solely to measurement errors.

\subsection{Comparison on different devices}

 \begin{figure}[t!]
	\centering
	\includegraphics[width=1\columnwidth]{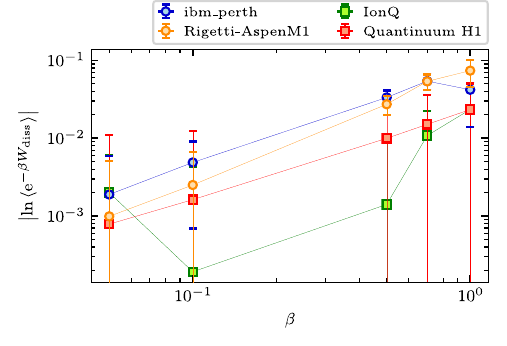}
    \caption{Validation of the Jarzynski equality Eq.~\eqref{eq:Jarzynski} for $L=4$ qubits as a function of inverse temperature $\beta$ on ibm\_perth~(blue), Rigetti Aspen-11~(orange) and IonQ~(green),  $65,536$ shots with classical sampling. Red: comparison with a circuit executed on Quantinuum H1 with a slightly different sampling, see text, and $4,000$ shots.
    The data for superconducting platforms are denoted by circles, for trapped ions -- by squares. The trapped ion platforms show a better performance than the superconducting qubit architectures, since $\ln \langle \e^{-\beta W_\text{diss}} \rangle$ is closest to zero -- the theory prediction value.
    }
	\label{fig:L4comp}
    \end{figure}

Besides presenting results for the largest system sizes and the most resource-intensive simulations (see main text), it is worth comparing the performance of the different devices on smaller systems. This is necessary to obtain an overview of the scaling with system size $L$ of the simulations, and their expected resource costs. Furthermore, it allows us to get valuable insights about the quality of the experimental platforms. 
    
For a direct comparison of the  performance on the five  different devices, we estimate the validity of Jarzynski's equality using $2^{16}$ measurement shots and $L=4,8$ qubits. In the case of Quantinuum, we are restricted by the smaller number of measurements to $L=4$ qubits and to a different sampling scheme due to the higher financial resource costs for circuit evaluation on these devices: We sample each eigenstate 200 times, and weigh the results by the Boltzmann distribution during classical postprocessing. In this way, we are able to extract data for different inverse temperatures with a total of $4,000$ shots. 
 
Because of hardware limitations on Rigetti and IonQ, that prevent the use of mid-circuit measurements, in order to make a fair comparison between the various devices, in the following we use classical presampling for the preparation of the Gibbs ensemble.
    
\begin{table}
\label{q8tabular}
\centering
\begin{tabular}{||c| c |r ||} 
 \hline
 \textbf{Device} & $\left\vert\ln\langle\e^{- \beta W_{\text{diss}}}\rangle_{P(W)}\right\vert$ & \textbf{Runtime} \\ [0.5ex] 
 \hline\hline
ibmq\_guadalupe & $0.05(4)$ & $51~\mu$s\\
\hline
IonQ & $0.03(4)$ & $\approx 20$ ms\\
\hline
Rigetti Aspen-11 & $0.04(3)$ & $\approx 18~\mu$s\\
\hline
 \rowcolor{Gray}
ibmq\_guadalupe (mid-circuit meas.) & $0.10(4)$ & $57~\mu$s\\
 \hline \hline
\end{tabular}
\caption{Accuracy of the Jarzynski equality for $\beta=0.7$, executed on different devices for an extension of the protocol from Fig.~\ref{fig:Thermalstate} to $L=8$ qubits. The case for mid-circuit measurements is shaded in grey. The values for the runtimes on IonQ and Rigetti are extrapolated, using information about the compiled circuit and gate times. The accuracy of the results on ibmq\_guadalupe is comparable to that on the trapped ion device IonQ.
}
\end{table}   

We emphasize that the comparison in this section is not a statement about the quality of the different devices for an accurate simulation of a quantum circuit: as explained in Sec.~\ref{sec:theory}, the quantum Jarzynski equality is only sensitive to noise channels violating double-stochasticity, and thus a good accuracy for the validation of the circuit does not imply an accurate evaluation of the corresponding quantum circuit.

\begin{table*}[t!]\centering
\label{table:technical}
\begin{tabular}{||c c c c c c c||} 
 \hline
 \textbf{Device name}  & \textbf{\# Qubits} & $T_1$[$\mu$s] & $T_2$[$\mu$s] & $T_g$[$\mu$s] & \textbf{Average two-qubit fidelity} & q\\ [0.5ex] 
 \hline\hline
 ibm\_perth~\cite{IBM} & 7 & 134 & 146 &0.44&0.988 &$3.0\cdot 10^2$\\ 
 \hline
 ibmq\_guadalupe~\cite{IBM} & 16 & 106& 119 & 0.4 &0.99&$2.6 \cdot 10^2$\\
 \hline
 IonQ~\cite{IonQ} & 11 & $10^{10}$ & $2\times 10^3$ &200&0.96&$5.0\cdot 10^7$\\
 \hline
  Rigetti Aspen-11~\cite{Rigetti} &40& 30 & 14 & 0.18 &$\sim 0.93$&$1.7\cdot 10^2$\\
 \hline
 quantinuum.hqs-lt-s1~\cite{quantinuum} & 20 & $>10^9$ & $3\times 10^6$ &28 &0.997&$>3\cdot 10^7$\\ [1ex] 
 \hline
\end{tabular}
\caption{Technical data of the various quantum devices.
$T_1$ is the relaxation rate of diagonal matrix elements in the density matrix, while $T_2$ denotes the relaxation rate of off-diagonal elements~\cite{nielsen2002quantum}.
$T_g$ denotes the average execution time of the native two-qubit gate on each of these devices. The average two-qubit fidelity is the average fidelity of the system-specific two-qubit gate. $q$ denotes the ratio between $T_1$ and gate time $T_g$, as introduced in App.~\ref{sec:Comparison}. While the qubit fidelity is comparable for all devices, $T_1$ varies over several orders of magnitude, depending on the underlying architecture.
}
\end{table*}

\textit{Accuracy of the results.---}The results for $L=4$ qubits are shown in  Fig.~\ref{fig:L4comp} for five different values of the inverse temperature $\beta$; for $L=8$ and a fixed $\beta=0.7$, see Table~\ref{q8tabular}. The error bars indicate the statistical uncertainty due to a finite number of measurements.
    
We apply no error mitigation; instead, we consider the measurement process itself as part of the protocol with its own errors, which, depending on their origin, may or may not violate the double stochasticity condition. Moreover, in the case of IonQ, error mitigation was infeasible, since the data collection continued over a few days. However, we can still obtain rough estimates for the measurement errors and their impact on the validity of Jarzynski's equality.
    
The errors for the IonQ, Rigetti and IBM devices are within 2\% to 5\%, even for large values of $\beta$ and $L=8$ qubits. Note that the theory predicted value from Eq.~\eqref{eq:Jarzynski}, $\ln{ \braket{\e^{- \beta W_{\text{diss}}}}_{P(W)}}=0$, does not fall within the error bars for the superconducting architectures---a direct manifestation of violations of double-stochasticity.

Even if deviations from Jarzynski's equality are not detectable with the current experiment, a closer look reveals that the measurement process itself is an energy-dissipative process in this case: our quantum simulations reveal that excited qubits are detected in the ground state with a probability of roughly 1\%. This process violates the second sum rule in Eq.~\eqref{eq:Sumrules}, and thus leads to a weak violation of the quantum Jarzynski equality. As discussed in App.~\ref{sec:Error}, our simulations reveal a means to detect the readout error as a residual deviation in the limit of infinite measurements.
    
It is also insightful to compare the deviation from Eq.~\eqref{eq:Jarzynski} with and without error-prone mid-circuit measurements on a IBM quantum device. In the case of $L=8$ qubits, the error is increased by almost a factor of 2 due to the first measurement, as is shown in Table~\ref{q8tabular}. This already shows that most of the deviation from the theoretical prediction is due to the measurement process.

On Rigetti's Aspen-11, we found it was essential to disable ``fencing'' to obtain quality results on par with other devices. Fencing makes two-qubit gates executed sequentially, even when acting on different qubits within the same circuit layer. While this reduces crosstalks and increases the fidelity of the individual operations, it leads to circuits with a longer execution time, making relaxation effects through $T_1$ more prominent.

\textit{Runtime and resource cost.---}Regarding the execution time, note that different physical platforms have different run times. In the case of superconducting qubits, gates are implemented as a sequence of microwave pulses, where the average duration of each such gate is of the order of $10$ ns to $100$ ns~\cite{Sheldon2016Procedure}; thus, the overall circuit duration for $L=8$ qubits is of the order of 50~$\mu$s. In the case of trapped ions, gates are implemented via two-photon Raman processes, with gate times ranging from $10$~$\mu$s to $100$~$\mu$s~\cite{IonQ}. As a consequence, the execution time for a circuit is also three orders of magnitude longer, which presents a relevant bottleneck for us when increasing the system size or the number of measurement shots.
    
The use of different architectures also affects the simulation cost. The simulation cost on trapped ion systems is at least 30 times higher than on superconducting qubits, which is caused by the longer circuit evaluation times. This is the major bottleneck we encounter for simulations on trapped ion devices: it limits the number of circuit evaluations to $2^{16}$ and the system size to $L=8$ qubits on IonQ, and to $4,000$ evaluations on $L=4$ qubit systems on Quantinuum H1, respectively. 
    
\textit{Ability to perform midcircuit measurements---}
In the end, we want to perform a quantum simulation without any classical presampling.
The thermal state preparation introduced in Sec.~\ref{subsec:Gibbs} requires the ability to perform midcircuit measurements. From the above platforms, only the IBM and Quantinuum devices are currently capable of performing this task.

\subsection{Technical data of the various quantum devices}\label{sec:Technical}

The technical details, including coherence times, thermalization times, and gate times, as well as the average two-qubit fidelities on the different devices we used, are shown in Table~\ref{table:technical}.

\section{Preparation of a thermal distribution for the transverse field Ising model}\label{Ising preparation}

In the following section, it is explained how the eigenstates of the transverse field Ising model are mapped to computational basis states of the quantum simulator by a shallow circuit. This step is crucial to test the quantum Jarzynski equality on current devices.

\subsection{Theory}

As we discussed in the main text, the transverse field Ising model can be mapped to a noninteracting fermionic Hamiltonian. It is therefore possible to prepare it in a Gibbs ensemble, using the protocol described in Sec.~\ref{subsec:Gibbs}, and with the additional help of a unitary transformation between the energy and spin bases of the system. The different transformation steps are explained here in detail, following Ref.~\onlinecite{CerveraLierta2018exactisingmodel}. 

The transverse field Ising model is given by the Hamiltonian
\begin{equation}
    H^\prime=\sum_{i=1}^L \sigma^x_{i} \sigma^x_{i+1} +\sum_{i=1}^L \sigma^z_i.
\end{equation}
The transverse field here is chosen of the same strength as the Ising interaction, although this is not a necessary requirement for testing Jarzynski's equality using our protocol.
In order to simplify the realization on a quantum computer, we impose periodic boundary conditions, and add an additional Pauli string to eliminate unwanted terms that appear in the Jordan-Wigner transformation: 
\begin{equation}
	H=\sum_{i=1}^L \sigma^x_{i} \sigma^x_{i+1} + \sum_{i=1}^L \sigma^z_i+\sigma^y_1 \sigma^z_2 \dots \sigma^z_{L-1} \sigma^y_L.
\end{equation}
Note that the multi-body term becomes negligible in the thermodynamic limit.

As a first step, we transform the Hamiltonian into fermionic modes using a Jordan-Wigner transform:
\begin{align}
    c_j=\left(\prod_{i<j} \sigma^z_i\right) \frac{\sigma_j^x +i \sigma^y_j}{2}, \quad c^\dagger_j= \frac{\sigma_j^x +- \sigma^y_j}{2} \left(\prod_{i<j} \sigma^z_i\right).
\end{align}
This gives the following fermionic Hamiltonian: 
\begin{equation}
    H= \sum_{i=1}^{L}\frac{1}{2}\left(c^\dagger_i c_{i+1}+c^\dagger_{i+1} c_{i}+c_i c_{i+1} +c^\dagger_{i+1}c^\dagger_{i}\right) + c^\dagger_i c^\dagger_{i+1}. 
\end{equation}
The wave function can be expressed as 
\begin{eqnarray}
    \ket{\psi}&=&\sum_{i_1,\dot i_L=0,1} \psi_{i_1,\dots, i_L} \ket{i_1 \dots i_L}\nonumber\\
    &=& \sum_{i_1,\dots i_L=0,1} \psi_{i_1,\dots, i_L}\bigl(c^\dagger_{1}\bigr)^{i_1} \dots \bigl(c^\dagger_{L}\bigr)^{i_L}\ket{\Omega_L}.
\end{eqnarray}
Here $\ket{\Omega_N}$ is the vacuum state, i.e., $c_i \ket{\Omega_N}=0$.
Note that the coefficients of the wave function do not change; thus, the Jordan-Wigner transformation does not add any additional gates to the quantum circuit. However, we have to keep track of fermionic signs when swapping fermionic modes.

The next step is to apply a Fourier transform:
\begin{equation}
    \tilde{c}_k^\dagger=\frac{1}{\sqrt{N}} \sum_{j=0}^{L-1} \e^{i \frac{2 \pi j}{L} k} c^\dagger_j,\quad k=-\frac{L}{2}+1 \dots ,\frac{L}{2}.
\end{equation}
The Fourier transform for $L=2^m$ ($m\in\mathbb{N}$) qubits can be implemented with a quantum circuit of depth $\log(L)$.
To see how, we split the Fourier transform into even and odd sites:
\begin{equation}
    \sum_{j=1}^L \e^{i \frac{2 \pi j}{L} k} c^\dagger_j
    =\sum_{j^\prime=0}^{\frac{L}{2}-1} \left( \e^{\frac{2 \pi i k}{L/2} j^\prime} c^\dagger_{2 j^\prime} +\e^{ \frac{2 \pi i k}{L}} \e^{ \frac{2 \pi i k}{L/2} j^\prime} c^\dagger_{2 j^\prime+1} \right).
\end{equation}
The two terms on the right-hand-side independently represent a Fourier transform for $L/2$ fermions.  The case $L=2^m$ is particularly appealing, since we can keep iterating this step until we end with a Fourier transform of only two fermions, which can be easily implemented using two-qubit gates.

\begin{figure}[t!]
	\centering
	\includegraphics[width=1\columnwidth]{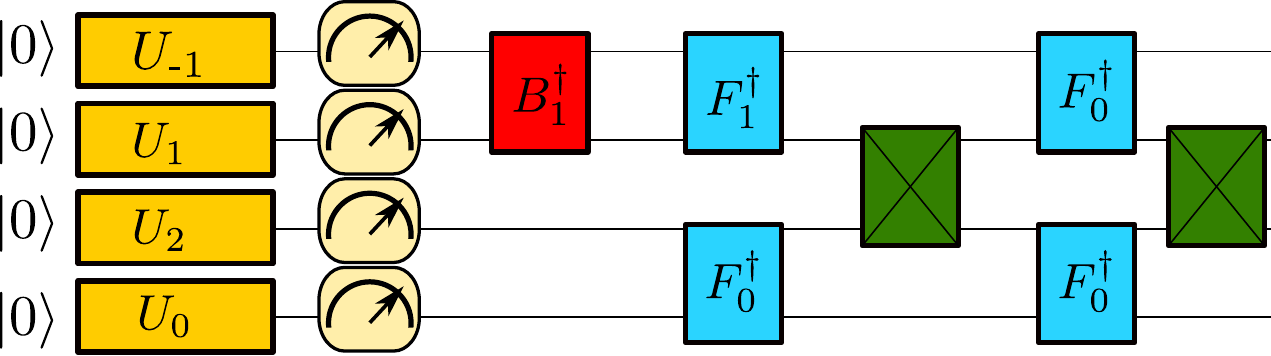}
	\caption{A circuit to prepare the thermal ensemble of the transverse Ising Hamiltonian. The unitaries $U_i$ are rotation gates defined in Eqs.~\eqref{rotation} and~\eqref{rotationequation}. The ensemble arising from repeated measurements is described by a Gibbs state. After the measurement, the circuit realizes a transformation from the energy eigenbasis of the transverse field Ising model to the computational basis. Here $F_i$ are Fourier gates, and $B_i$ -- Boguliobov gates. Note that fermionic SWAP gates are required, that contain an extra sign compared to qubit swap gates.}
	\label{fig:ThermalstateIsing}
\end{figure}

To do so, we need the fermionic swap gate (note the sign structure, green gates in Fig.~\ref{fig:ThermalstateIsing})
\begin{equation}
    fSWAP=\begin{pmatrix}
    1 &0&0 &0 \\
    0 &0&1 &0 \\
    0 &1&0 &0 \\
    0 &0&0 &-1 \\
    \end{pmatrix},
\end{equation}
and the Fourier gates (light blue gates in Fig.~\ref{fig:ThermalstateIsing})
\begin{equation}
    F_k=\begin{pmatrix}
    1 &0&0 &0 \\
    0 &\frac{1}{\sqrt{2}}&\frac{\e^{\frac{2 \pi i k}{L}}}{\sqrt{2}} &0 \\
    0 &\frac{1}{\sqrt{2}}&\frac{-\e^{\frac{2 \pi i k}{L}}}{\sqrt{2}} &0 \\
    0 &0&0 &-\e^{\frac{2 \pi i k}{L}} \\
    \end{pmatrix}.
\end{equation}
In our case, we have to restrict to $L\leq16$, due to a limited number of available qubits.

\begin{figure}[t!]
	\centering
	\includegraphics[width=\columnwidth]{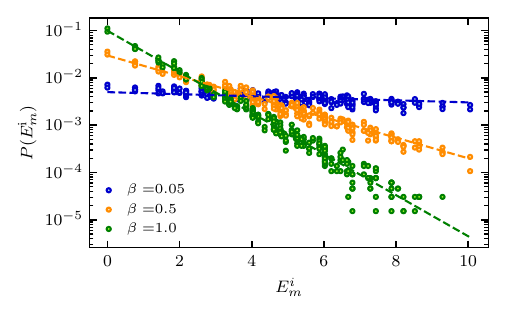}
	\caption{Probability $P(E^{\mathrm i}_m)$ to measure a given eigenstate $H_{\mathrm i} \ket{m^{\mathrm i}}= E^\mathrm i_m \ket{m^{\mathrm i}}$ as a function of energy $E^{\mathrm i}_m$, quantum simulations (filled dots) vs.~theory prediction $\mathrm{e}^{-\beta E^{\mathrm i}_m}/Z_{\mathrm i}$~(dashed lines), for different inverse temperatures $\beta$, at $L=8$. The simulation prepares up to small fluctuations the correct thermal state.
		}
	\label{fig:Temperaturetest}
\end{figure}

The above transformations lead to the Hamiltonian 
\begin{align}
\begin{split}
    H=\sum_{k=-L/2+1}^{L/2}\left[1-\cos\left( \frac{2 \pi k}{L}\right)\right] \tilde{c}^\dagger_k  \tilde{c}_k  \\ -i \sin \left( \frac{2 \pi k}{L}\right) (\tilde{c}^\dagger_{-k}\tilde{c}^\dagger_{k}+\tilde{c}_{-k}\tilde{c}_{k}),
\end{split}
\end{align}
Finally, in order to diagonalize the Hamiltonian, we have to apply a Bogoliubov transformation:
\begin{align}
    a_k=u_k \tilde{c}_k+ iv_k \tilde{c}^\dagger_{-k}, \nonumber \\
    a_k^\dagger=u_k \tilde{c}_k- iv_k \tilde{c}^\dagger_{-k}.
\end{align}
This transformation can be achieved with gates~(red gate in Fig.~\ref{fig:ThermalstateIsing}) of the form 
\begin{equation}
    B_k=\begin{pmatrix} \cos\frac{\phi_k}{2} & 0& 0 &i \sin\frac{\phi_k}{2}\\
    0 &1 &0 &0 \\
    0& 0 & 1 &0\\
    i \sin\frac{\phi_k}{2} & 0& 0 &i \cos\frac{\phi_k}{2}\end{pmatrix},
\end{equation}
where~\cite{CerveraLierta2018exactisingmodel}
\begin{equation}
    \phi_k=\arccos\left( \frac{1 -\cos\left(\frac{2 \pi k}{L}\right)}{\sqrt{\left[1-\cos\left(\frac{2 \pi k}{L}\right)\right]^2+\sin^2\left(\frac{2 \pi k}{L}\right)}}\right).
\end{equation}

The transformation steps for $L\geq4$ qubits are analogous to the case described above.
This casts the Hamiltonian in diagonal form: 
\begin{equation}
    H=\sum_{k=-L/2+1}^{L/2} \omega_k a^\dagger_k a_k +E_c,
\end{equation}
with eigenenergies 
\begin{equation}\label{wks}
    \omega_k=\sqrt{\left[1-\cos\left(\frac{2 \pi k}{L}\right)\right]^2+\sin^2\left(\frac{2 \pi k}{L}\right)}.
\end{equation}
Here, $E_c$  is a constant energy offset $E_c=1-\sqrt{2}$. Note that we ignore this term in the work distributions and free energy computations, since it only appears as a constant factor in the partition function $Z_\mathrm i$, which is not relevant for testing the validity of Eq.~\eqref{eq:Jarzynski}.

In order to prepare the Gibbs state, we consider the diagonalized Hamiltonian. In this case a thermal ensemble can be prepared using projective measurements, as explained in Sec.~\ref{subsec:Gibbs}. To apply a transformation back into the computational basis, we have to reverse the unitary Fourier and Bogoliubov transformations described above. The corresponding circuit is shown in Fig.~\ref{fig:ThermalstateIsing}.

\begin{figure}[t!]
	\centering
	\includegraphics[width=1\columnwidth]{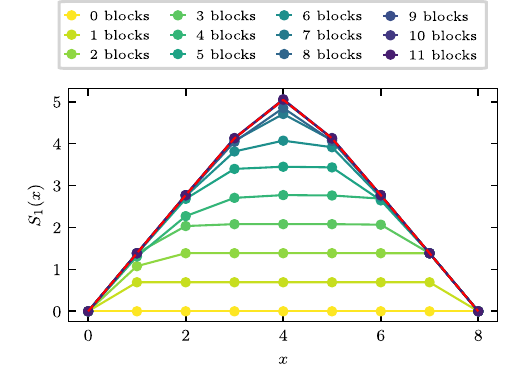}
	\caption{Classically computed operator entanglement entropy $S_1(x)$, Eq.~\eqref{eq:defentanglement} of the circuits as a function of the cut-position x, for a different number of blocks of the non-equilibrium protocol (legend), for $L=8$ qubits. We chose $L-1$ blocks for our simulations, see Fig.~\ref{fig:Thermalstate}.
 The dashed horizontal lines indicate the Page curve $S_\mathrm{Page}=2 z\ln 2 - 2^{2z-L-1}$, with $z=\min(z,L-z)$. $L-1$ blocks are sufficient to generate an entanglement curve close to the Page curve.}
	\label{fig:operatorentanglement}
\end{figure}

\subsection{Accuracy of the thermal state preparation using midcircuit measurement}

Let us now investigate the accuracy of the mid-circuit measurement state preparation.
To do so, we detect the probability to prepare a state $\ket{m^\mathrm{i}}$ after part (A) of the protocol displayed in Fig.~\ref{fig:Thermalstate}, and compare it with the probability distribution of the canonical ensemble. The results are shown in Fig.~\ref{fig:Temperaturetest}, using the data for $L=8$ qubits from Fig.~\ref{fig:L16}. The quantum simulation prepares the correct thermal ensemble for different inverse temperatures $\beta$ and the entire range of initial energies. Note that we observe a slight temperature dependence in the fluctuations: Low energy states are prepared more often than predicted by theory; higher energy states are slightly underrepresented. As we discussed in Sec.~\ref{sec:theory_analysis}, the measurement itself is a dissipative process, causing decay from the excited to the ground state of the qubits.

In general, device imperfections do not allow us to perfectly prepare a given target state. In our case, we obtain instead of an eigenstate $\ket{\phi_i}$ of the transverse field Ising model, a density matrix $\rho_{\text{sim}}^i$. To determine the quality of the state preparation process, we obtain the form of the density matrix for $L=4$ qubits using quantum state tomography ~\cite{Vogel1989Determination}.
This allows us to compute the average single particle fidelity
\begin{equation}
  F=\sum_{i=0}^{2^L}\sqrt[L]{\tr\{\rho_{\text{sim}^i} \ket{\phi_i}\bra{\phi_i}}\} \approx 0.59. 
\end{equation}
 As before we apply no error mitigation in this case.

\section{Analysis of the nonequilibrium protocols }\label{sec:Entanglement}

This section is dedicated to an analysis of the specific circuit protocols we selected to use in this study. We show here numerically that we are operating in a nontrivial quantum many-body regime, by computing the operator entanglement entropy of our circuits~\cite{zhou_operator_2017}, and the work distribution they give rise to. Finally, we present results for different choices of the one-body random unitary gates and show that the specific choice of random gates has only a minor impact on the accuracy for validation of Eq.~\eqref{eq:Jarzynski}. 

\begin{table}[t!]
\label{Table work}
\centering
\begin{tabular}{||c c c||} 
 \hline
 $\beta$ &  $\ln{ \braket{\e^{- \beta W_{\text{diss}}}}}$& $\braket{\beta W_{\text{diss}}}$  \\ [0.5ex] 
 \hline\hline
 0.1 & 0.01 & 0.004 \\ 
 \hline
 0.5 & 0.08 & 0.27  \\
 \hline
 1.0 & 0.23 & 1.03  \\
 \hline
\end{tabular}
\caption{$\braket{\e^{- \beta W_{\text{diss}}}}$ and $\beta W_{\text{diss}}$, experimental data for $L=8$ qubits. The data indicate that our chosen protocol is away from the adiabatic regime, for which $\langle W_\mathrm{diss}\rangle{=}0$.}
\end{table}

\subsection{Operator entanglement entropy}

We analyze our circuits from an entanglement perspective and show that our ideal circuits are sufficient to create entanglement close to the Page value. This, in turn, demonstrates that we operate in the quantum many-body regime. 

For a given system, we can choose complete basis sets of operators $\{ A_i\}$ and $\{B_i\}$ which are orthonormal and have only support on subsystem $A$ and its complement $B=A^c$, respectively.
An operator $O$ can now be decomposed as
\begin{equation}
    O=\sum_{i,j} O_{i,j} A_i \otimes B_j.
\end{equation}
This allows us to define the notion of a reduced operator density matrix $\rho^A_\mathrm{op}$ with matrix elements
\begin{equation}
    \left(\rho^A_\mathrm{op}\right)_{i,j}=\sum_{k} O_{i,k} O^*_{j,k} .
\end{equation}
In the following, we consider subsets of the form $A=\{0,\dots x\}$, where $x$ is the position of the last qubit included in the subset.
The operator entanglement entropy for such a partition is defined as~\cite{zhou_operator_2017}
\begin{equation}\label{eq:defentanglement}
    S_1(x)=-\tr\left(\rho^A_\mathrm{op} \ln \rho^A_\mathrm{op}\right)|_{A=\{0,\dots x\}}
\end{equation}

We now compute the operator entanglement entropy of the unitary operator of our protocol for different partitions.
 The results are shown in Fig.~\ref{fig:operatorentanglement}. It is clear that our chosen operators with $L-1$ blocks already exhibit an entanglement (operator) entropy close to the Page-value~\cite{Page1993Average},  $S_\mathrm{Page}=L\ln{2} -{1}/2$.

\subsection{Work distribution}

\begin{figure}[t!]
	\centering
	\includegraphics[width=1\columnwidth]{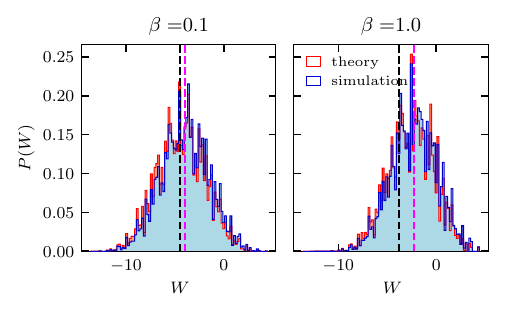}
	\caption{Histograms of the work distribution $P_F(W)$ for the protocol in Fig.~\ref{fig:Thermalstate} with seven blocks and $L=8$ qubits: theory prediction~(red) vs.~experimental simulations~(blue).The free energy difference $\Delta F$ is indicated by a dashed black line, the average work by a magenta line. Although the measured work can be smaller than the free energy for some shots, the average work is larger than the free energy difference and the second law of thermodynamics thus holds.
	The work distributions have a nearly Gaussian shape. The absence of strong tails in the work distribution reduces the number of required measurements.
	}
	\label{fig:Workdistribution}
\end{figure}

The work distributions $P_F(W)$ for two different inverse temperatures $\beta=0.1$ and $\beta=1.0$ are shown in Fig.~\ref{fig:Workdistribution}, for the protocol from Fig.~\ref{fig:Thermalstate}. We obtain an approximately Gaussian distribution for both temperatures. 

At this point it is interesting to compare the work distribution with the free energy difference. The free energy difference is marked in Fig.~\ref{fig:Workdistribution} by a dashed black line, the average work by a dashed magenta line. Since $\Delta F< \braket{W}_{P_F}$, the second law of thermodynamics holds, as expected. However, it is also visible that with a finite probability the extracted work $\braket{W}_{P_F}$ for single realizations is smaller than the free energy difference; such datapoints are known as "microscopic violations" of the second law~\cite{Maillet2019optimal}.

Furthermore, in order to demonstrate that our circuit operates away from the adiabatic regime, we compute $\braket{\beta W_{\text{diss}}}$ and compare it to Eq.~\eqref{eq:Jarzynski}, see Table~\ref{Table work}.
Since $W_{\text{diss}}\gg 0$, it follows that the adiabatic approximation does not hold for our chosen protocols.

\subsection{Different circuit realizations}

\begin{figure}[t!]
	\centering
	\includegraphics[width=1\columnwidth]{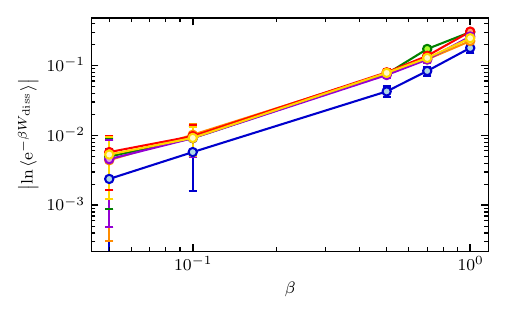}
	\caption{Validation of the Jarzynski equality for $L=4$ qubits as a function of inverse temperature $\beta$ for different choices of random unitaries in the nonequilibrium protocol, simulated on ibm\_perth. The accuracy of the result is in all cases comparable, i.e., it does not depend strongly on the choice of the random single-qubit gates in the protocol of Fig.~\ref{fig:Thermalstate}. 
	}
	\label{fig:Differentexperiments}
\end{figure}

Let us also check the effect of different single-qubit random unitaries on our results. To do so, we repeat our experiment for 4 qubits and 3 layers on ibm\_perth for different circuit realizations, by choosing different random unitary gates. As we can see in Fig.~\ref{fig:Differentexperiments}, the deviation from Jarzynski's equality does not depend strongly on the particular choice of random gates in our circuits. 

\subsection{Dependence on the number of circuit blocks}\label{subsec:blockdependence}

\begin{figure}[t!]
	\centering
	\includegraphics[width=1\columnwidth]{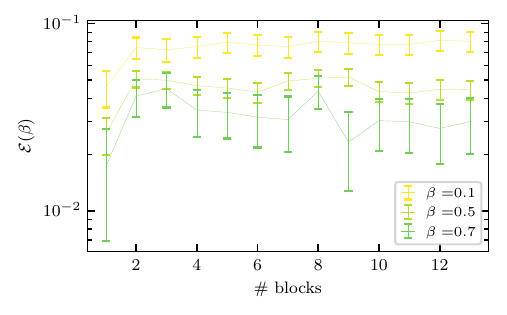}
	\caption{ Test of the Jarzynski equality Eq.~\eqref{eq:Jarzynski} as a function of the circuit depth of the non-equilibrium protocol. The legend shows different inverse temperatures $\beta$ for a system of $L=8$ qubits. A single block here is defined in part (C) of the circuit introduced in Fig.~\ref{fig:Thermalstate}. The deviations from~Eq.~\eqref{eq:Jarzynski} are almost independent of the number of blocks in the non-equilibrium protocol.
	}
	\label{fig:Blocksize}
\end{figure}

Finally, we investigate the deviations from Eq.~\eqref{eq:Jarzynski} as a function of the non-equilibrium-protocol length. To do this, we consider the blocks introduced in part (C) of the circuit in Fig.~\ref{fig:Thermalstate}; we can then stack multiple such blocks with different unitaries one after the other to increase the circuit depth. As is shown in Fig.~\ref{fig:Blocksize}, the deviations barely scale with the size of the non-equilibrium protocol for more than two-three blocks. In this case, most of the errors accumulated during the circuit are compensated for by applications of single-qubit random gates, as is explained in Sec.~\ref{sec:theory_analysis}.

\section{\label{sec:Error}Statistical fluctuations due to a finite number of measurements}

The following section gives a quantitative analysis of the statistical uncertainties in the case of a finite number of measurements. 

\subsection{General theory}

For the scrambling circuits we choose in our simulations, we can use additional assumptions to get simple estimates for the size of the statistical fluctuations. This is especially helpful if we want to estimate the number of shots needed to estimate the free energy difference to a given accuracy.

To do so, consider the quantity
\begin{equation}\label{eq:dev}
    \Delta P=\Bigl\langle\e^{-\beta(\Delta W-\Delta F)}\Bigr\rangle.
\end{equation}
This is just another representation of Jarzynski's equality from Eq.~\eqref{eq:Jarzynski}. Given a perfect experimental realization including an infinite number of measurements, we expect $\Delta P=1$. All deviations from this value can therefore be assigned to statistical errors, or errors coming from noise violating double-stochasticity. Note that we consider $\Delta P$ instead of Eq.~\eqref{eq:Jarzynski}, since this representation is easier to handle analytically. Furthermore, in the limit of small deviations, the first order Taylor expansions around unity for $\Delta P$, and zero for Eq.~\eqref{eq:Jarzynski}, agree with one another. 

Given the eigenenergies of the initial and final Hamiltonians, we compute the contribution $\e^{\beta \Delta F}={Z_\mathrm{i}}/{Z_\mathrm{f}}$ exactly. Thus, any error originates from the measurement of the work distribution. 
We write
\begin{equation} \label{eq:Jarwitherrors}
    \Bigl\langle\e^{-\beta W}\Bigr\rangle=\sum_{m,n} P_m K_{m\rightarrow n} \e^{-\beta(E^\mathrm{i}_m-E^\mathrm{f}_n)}. 
\end{equation}
Here $P_m$ denotes  the probability to prepare a given initial eigenstate, and $K_{m\rightarrow n}$ is the transition matrix of the process, defined in Sec.~\ref{sec:theory}. In the optimal case of infinitely many measurements and error-free  preparation, we have $P_m=\e^{-\beta E^\mathrm i_m}/{Z_\mathrm{i}}$.

There are two generic ways for errors to occur.
The first one is as a statistical error in $P_m$, 
\begin{equation}
    P_m=\frac{1}{Z_\mathrm{i}} \e^{-\beta E^\mathrm{i}_m} +\delta P_m.
\end{equation}
The first term on the right-hand side in the above equation is the probability distribution of the canonical ensemble. The second term denotes the statistical error for the measurement probability of each state $\ket{m^\mathrm{i}}$, which satisfies the sum rule $\sum_m \delta P_m=0$. For a large enough number of states, we can assume the errors $P_m$ to be independent of one another and neglect this constraint. 
The statistical error of measurement probability for each state $\ket{m^\mathrm{i}}$ can be modeled by a binomial distribution: For each shot, we obtain this state with probability $\mathrm e^{-\beta E^\mathrm{i}_m} /Z_\mathrm{i}$ and measure another state with probability $1-\mathrm e^{-\beta E^\mathrm{i}_m} /Z_\mathrm{i}$; then the statistical uncertainty is given by the variance of the binomial distribution, and scales as 
\begin{equation}
    \delta P_m \propto \sqrt{\frac{\e^{-\beta E^\mathrm{i}_m}}{Z_\mathrm{i}} \left(1-\frac{\e^{-\beta E^\mathrm{i}_m}}{Z_\mathrm{i}}\right) \frac{1}{s}},
\end{equation}
where $s$ denotes the number of shots. This expression is the variance of a binomial distribution for an event occurring with probability ${\e^{-\beta E^\mathrm{i}_m}}/{Z_\mathrm{i}}$.

The second way an error can occur is through a measurement of $K_{m\rightarrow n}$.
We can write
\begin{equation}
    K_{m\rightarrow n}=\tilde{K}_{m\rightarrow n}+\delta K_{m\rightarrow n},
\end{equation}
where $\tilde{K}_{m\rightarrow n}$ comprises any unitary or doubly-stochastic contributions that Jarzynski's equality is insensitive to. Using the (simplifying) assumption that the correct size of each matrix element is roughly ${1}/{D}$ with $D=2^L$ (this is justified by using a scrambling circuit, and becomes more accurate with increasing system size $L\gg 1$), the error of $\delta K_{m\rightarrow n}$ scales with the number of measurement shots as
\begin{equation}
    \delta K_{m\rightarrow n} \propto \sqrt{\frac {Z_\mathrm{i} }{D s \e^{-\beta E^\mathrm{i}_m}}}.
\end{equation}
This is again the standard deviation for a binomial process of an event with probability ${1}/{D}$ and an effective number of
repetitions ${s \e^{-\beta {E^\mathrm{i}_m}}}/{Z_\mathrm{i}}$. We used here $(1-{1}/{D})\approx 1$.
\begin{figure}[t!]
	\centering
	\includegraphics[width=1\columnwidth]{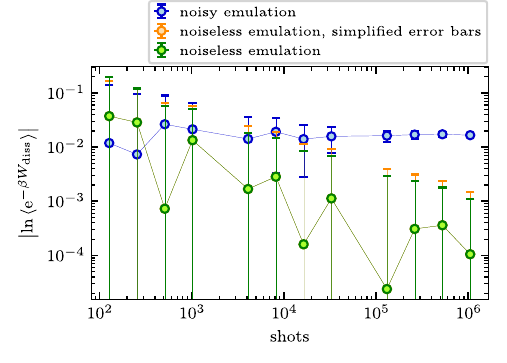}
	\caption{Classically emulated validation of the Jarzynski equality Eq.~\eqref{eq:Jarzynski} for $L=8$ qubits as a function of the number of shots, noise-free results at inverse temperature $\beta=0.1$. The green error bars indicate the error due to a finite number of measurements in the emulation. The orange error bars are computed using the simplification of roughly equal-size transition matrix elements [see text].  Blue: Classical emulation, but now with thermal noise of similar size as on ibmq\_guadalupe.
	The impact of shot noise becomes negligible with an increasing number of measurements (error bars decrease), such that the effects of violating double-stochasticity become visible (remaining finite plateau value).
	}
	\label{fig:Errorscaling}
\end{figure}

Taking these considerations into account, we can now divide the contributions of the fluctuations into four parts:
\begin{align}
    \Delta P&=1 \label{theoryterm}\\&+\frac{Z_\mathrm i}{Z_\mathrm f}\sum_{m,n} \delta P_m \tilde{K}_{m\rightarrow n} \e^{-\beta(E^\mathrm{i}_m-E^\mathrm{f}_n)} \label{cont1}\\&+
    \frac{1}{Z_\mathrm f} \sum_{m,n} \delta K_{m\rightarrow n} \e^{-\beta E^\mathrm{f}_n} \label{cont2}
    \\&+\frac{Z_\mathrm i}{Z_\mathrm f} \sum_{m,n} \delta P_m\delta K_{m\rightarrow n} \e^{-\beta(E^\mathrm{i}_m-E^\mathrm{f}_n)} \label{cont3}.
\end{align}
Equation~\eqref{theoryterm} is Jarzynski's equality.
Assuming that all terms $K_{m\rightarrow n} \sim {1}/{N}$ (as a consequence of using random circuits, which should not prefer any particular eigenstate transitions), and invoking the central limit theorem (the variance of a sum is the sum of the variances), we find that the second term, Eq.~\eqref{cont1}, scales as 
\begin{align}
    \eqref{cont1}\propto \frac{Z_\mathrm{i}}{D\sqrt{s}}\sqrt{\sum_{_m} \frac{\e^{\beta E^\mathrm{i}_m}}{Z_\mathrm{i}} \left(1-\frac{\e^{-\beta E^\mathrm{i}_m}}{Z_\mathrm{i}}\right) }.
\end{align}
In turn, the third term is proportional to
\begin{align}
    \eqref{cont2}\propto \sqrt{\frac{Z_\mathrm{i}}{D s}} \sqrt{\sum_{m}\frac{1}{\e^{-\beta E^\mathrm{i}_m}}} \sqrt{\sum_ {n}\frac{\e^{-2 \beta E^\mathrm{f}_n}}{Z_\mathrm{f}^2}},
\end{align}
and the last term is of order 
\begin{align}
    \eqref{cont3}\propto \frac{Z_\mathrm{i}}{\sqrt{N} s Z_\mathrm{f}} \sqrt{\sum_{m} \e^{2 \beta E^\mathrm{i}_m}\left(1-\frac{\e^{-\beta E^\mathrm{i}_m}}{Z_\mathrm{i}}\right)} \sqrt{\sum_ {n}\e^{-2 \beta E^\mathrm{f}_n}}.
\end{align}

Let us now test these expressions and the underlying assumptions we made above. Figure~\ref{fig:Errorscaling} shows the scaling of the errors for a noise-free quantum simulation~(red) and a comparison with the exact error bars using the exact size of the matrix elements~(blue). The data show that the simplification we used to compute error bars is well justified.

\subsection{High- and low-temperature expansions for deviations from Jarzynski's equality}\label{sec: high- and low temperature expansion}

The goal of this section is to explain 
the dependence of deviations from Eq.~\eqref{eq:Jarzynski} in quantum simulations, as is shown in Fig.~\ref{fig:L16}. In order to make the discussion more transparent, we ignore errors in the preparation of the initial state and due to a finite number of measurements and focus here on the violations of double-stochasticity instead. To be more concrete, we only consider errors of type Eq.~\eqref{cont2}.

\textbf{High-temperature expansion:}
The conservation of probability gives
\begin{equation}\label{eq:sumrule}
     \sum_{m,n}\delta K_{m\rightarrow n}=0.
\end{equation}
Furthermore, if we denote by $D$ the Hilbert space dimension, a Taylor expansion for small $\beta$ gives
\begin{align}
    \Delta P&=1 
    + \frac{\beta}{D} \sum_{m,n} \delta K_{m\rightarrow n} E^\mathrm{f}_n +\mathcal{O}(\beta^2). 
    \label{eq:small-invtemperatures}
\end{align}
We note that due to Eq.~\eqref{eq:sumrule} this expression is insensitive to constant shifts of the overall energy.

For high temperatures, the deviation scales linearly with inverse temperature $\beta$. Transitions towards all states, therefore, give rise to maximum deviation.

\textbf{Low temperature expansion:}
The behavior drastically changes, if one instead considers inverse temperatures larger than the inverse gap to the ground state energy $E^\mathrm{f}_0$.
In this case, $Z_\mathrm f\approx e^{-\beta E^\mathrm{f}_0}$ and only transitions to the ground state contribute:
\begin{align}
    \Delta P&\approx 1 
    + \sum_{m} \delta K_{m\rightarrow 0}  \label{eq:small-temperatures}.
\end{align}
The deviation converges in this case towards a constant value, determined by the violations of double-stochasticity w.r.t.~the ground state.

\begin{figure}[t!]
	\centering
	\includegraphics[width=1\columnwidth]{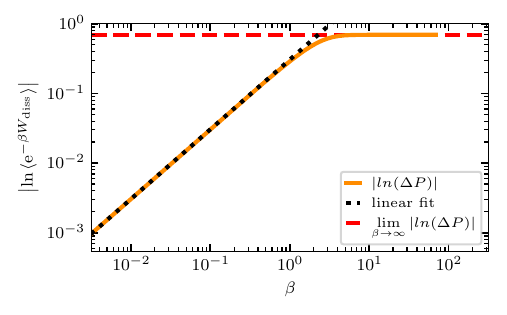}
	\caption{$|ln(\Delta P)|$~Eq.~\eqref{eq:dev} as a function of $\beta$~(orange), together with the high-temperature expansion Eq.~\eqref{eq:small-invtemperatures}~(dotted black line) and the zero-temperature limit Eq.~\eqref{eq:small-temperatures}~(red). The transition matrix is extracted from simulations data with $L=8$ and $\beta=0.05$.
    The deviations from Eq.~\eqref{eq:Jarzynski} for a dissipation-free simulations agree perfectly with the high- and low-temperature expansions (see text). 
	}
	\label{fig:Errorscalingbeta}
\end{figure}

\textbf{Comparison with simulations:}
To test the approximations above, we extract the transition matrix from data for $\beta=0.05$ and $L=8$. Afterwards, we weigh all transitions using the weights of the canonical ensemble which allows us to construct curves for all values of $\beta$. The results are shown in Fig.~\ref{fig:Errorscalingbeta}:
the deviations from Eq.~\eqref{eq:Jarzynski} for dissipation-free simulations agree perfectly with the high- and low-temperature expansion described above. 

\section{Relation between $T_1$ measurement and the Jarzynski equality.}\label{sec:T_1}

In the following section, we illustrate the connection between the standard measurement scheme for $T_1$~\cite{nielsen2002quantum}
and the quantum Jarzynski equality. This derivation also illustrates how the latter can be interpreted as a generalization of all experimental protocols detecting processes violating double-stochasticity like energy dissipation.

Consider a single-particle system described by the Hamiltonian
\begin{align} \label{eq: single particle Hamiltonian}
H_\mathrm{i} =H_\mathrm{f}=E \left(\ket{1}\bra{1} - \ket{0}\bra{0}\right).
\end{align}
The specific choice of the energies does not matter, as long as they are non-degenerate.  

We can interpret the standard protocol to measure the thermal relaxation time $T_1$ as a dynamical process connecting the initial and final ensemble:
\begin{align*}
    \includegraphics[width=.25\textwidth]{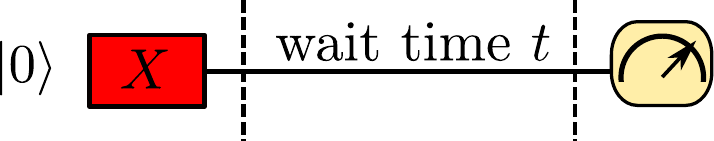}
\end{align*}
First apply a $X$ gate to excite the qubit, followed by some waiting time $t$. 
The fraction of decayed states is thus given by
\begin{align}\label{eq:population}
    p_1=1-\mathrm{e}^{-\frac{t}{T_1}}.
\end{align}

Similar to the analysis presented in the main text, $\ket{\psi_G}=(1,0)$ defines the "physical" ground state of the qubit, whereas $\ket{\psi_E}=(0,1)$ denotes an excited state.

In order to simplify our analysis, we assume that the state preparation is perfect and statistical errors due to a finite number of measurements are negligible. Furthermore, we assume that there are no unwanted excitations during the process from the ground state to the excited state, i.e., $\delta K_{0\rightarrow 1}=0$. 

To determine the deviation from the Jarzynski equality~Eq.~\eqref{theoryterm}, we have to determine the transition matrix $K_{m\rightarrow n}$, which is given by
\begin{align}\label{eq:matrix}
    K_{m\rightarrow n}=\begin{pmatrix} 1 &p_1\\
    0 &1-p_1  \\
\end{pmatrix}.
\end{align}

The assumptions above are sufficient to determine the deviation terms in Eq.~\eqref{theoryterm}-\eqref{cont3}:
\begin{eqnarray}
\label{eq:deviationsexample}
    \delta K_{1\rightarrow0} &=&-\delta K_{1\rightarrow 1}=p_1,\\
      \delta K_{0\rightarrow1}&=&\delta K_{0\rightarrow 0}=0,\\
    \delta P_0&=&\delta P_1=0.
\end{eqnarray}
The first two relations follow from Eq.~\eqref{eq:matrix}, the third is a consequence of perfect state preparation.

The evaluation of Eq.~\eqref{theoryterm}-\eqref{cont3} gives, together with Eq.~\eqref{eq: single particle Hamiltonian} and Eq.~\eqref{eq:deviationsexample},
\begin{align}
    \Delta P=1+p_1\tanh{\beta E}.
\end{align}
The deviation $p_1\tanh{\beta E}$ is proportional to the decayed population. This means that the standard protocol to measure $T_1$ is a special case of testing deviations from the quantum Jarzynski equality.

\section{Comparison with previous experiments}\label{sec:comparison}

In this appendix, we discuss the details of the comparison of our results to previous work, cf.~Table~\ref{Expcomparison}.

We use the definition from Eq.~\eqref{Eq:normalized} to compare our simulation with previous experiments on systems with a finite Hilbert space dimension. To do so, we extract the Hamiltonian, the inverse temperature $\beta$, and the measurement data for the validation of Eq.~\eqref{eq:Jarzynski} from the simulations. Using these data, the evaluation of Eq.~\eqref{Eq:normalized} is straightforward.

In the experiments~\cite{bassman2021computing,smith2018verification,Bartalhao2014Experimental,Solfanelli2021Experimental}, the underlying Hamiltonian was either a spin-$1/2$ system or the transverse field Ising model with up to three qubits. We extracted the data for the validation from Fig.~3 of Ref.~\cite{bassman2021computing}, from Fig.~3 of Ref.~\cite{smith2018verification}, from Fig.~3 of Ref.~\cite{Bartalhao2014Experimental}, Fig.~3 of Ref.~\cite{Solfanelli2021Experimental} and Table 1 of Ref.~\cite{cerisola2017using}. The case of Ref~\cite{hernandez2021experimental} is special: in their setup the authors chose $\beta=0$, where Jarzynski's equality is trivially obeyed, as can be directly seen from Eq.~\eqref{eq:Jarzynski}.

The trapped ion experiment~\cite{an2015experimental} was modeled by a harmonic oscillator, i.e., a system with an infinite-dimensional Hilbert space. In order to compare with the other experiments that deal with systems with a finite-dimensional Hilbert space, we evaluate Eq.~\eqref{Eq:normalized} and introduce an artificial cutoff for the harmonic oscillator Hilbert space by taking only the first 10 modes into account; the experimental data concerning the occupation of the different modes shows that this is a valid approximation. We normalized $\beta$ by the energy gap of the harmonic oscillator. For the extraction of the data, we considered Table 1 of Ref.~\cite{an2015experimental}, using a ramp time of $5$~$\mu$s.

\bibliography{bibi}

\begin{thebibliography}{75}
\bibitem [{Note1({\natexlab{b}})}]{Notefig4}%
  \BibitemOpen
  \bibinfo {note} {It is important to avoid confusion about the physical units
  of $T_1$; it denotes an intrinsic timescale of our physical device. In the
  analysis of the Jarzynski equality, $T_1$ enters indirectly, and only via the
  ratio of gate time $T_g$ and $T_1$, which is dimensionless
  again.}\BibitemShut {Stop}%
\end{thebibliography}

\end{document}